\documentclass[twoside,11pt]{article}

\usepackage[utf8]{inputenc}
\usepackage{CJKutf8}
\DeclareUnicodeCharacter{1EB9}{\d{e}}

\usepackage{jair}
\usepackage{multirow}
\usepackage{float}
\usepackage{array}
\usepackage{csquotes}
\usepackage{amsmath}
\usepackage{graphicx}
\usepackage{amssymb}
\usepackage{pifont}
\usepackage[caption=false]{subfig}
\usepackage[table]{xcolor}
\usepackage{url}
\usepackage{hyperref}

\newcommand{\xmark}{\ding{53}}%

\begin{document}

\title{Assessing Gender Bias in Machine Translation -- A Case Study with Google Translate}

\author{\name Marcelo Prates \email morprates@inf.ufrgs.br \\
\name Pedro Avelar \email pedro.avelar@inf.ufrgs.br \\
\name Luis C. Lamb \email lamb@inf.ufrgs.br \\
\addr Federal University of Rio Grande do Sul}

\maketitle

\begin{abstract}
Recently there has been a growing concern in academia, industrial research labs and the mainstream commercial media about the phenomenon dubbed as \emph{machine bias}, where trained statistical models -- unbeknownst to their creators -- grow to reflect controversial societal asymmetries, such as gender or racial bias. A significant number of Artificial Intelligence tools have recently been suggested to be harmfully biased towards some minority, with reports of racist criminal behavior predictors, Apple's Iphone X failing to differentiate between two distinct Asian people and the now infamous case of Google photos' mistakenly classifying black people as gorillas. Although a systematic study of such biases can be difficult, we believe that automated translation tools can be exploited through gender neutral languages to yield a window into the phenomenon of gender bias in AI.

In this paper,  we start with a comprehensive list of job positions from the U.S. Bureau of Labor Statistics (BLS) and used it in order to build sentences in constructions like ``He/She is an Engineer'' (where ``Engineer'' is replaced by the job position of interest) in 12 different gender neutral languages such as Hungarian, Chinese, Yoruba, and several others. We translate these sentences into English using the Google Translate API, and collect statistics about the frequency of female, male and gender-neutral pronouns in the translated output. We then show that Google Translate exhibits a strong tendency towards male defaults,  in particular for fields typically associated to unbalanced gender distribution or  stereotypes such as STEM (Science, Technology, Engineering and Mathematics) jobs. We ran these statistics against BLS' data for the frequency of female participation in each job position, in which we show that Google Translate fails to reproduce a real-world distribution of female workers. In summary, we provide experimental evidence that even if one does not expect in principle a 50:50 pronominal gender distribution, Google Translate yields male defaults much more frequently than what would be expected from demographic data alone.

We believe that our study can shed further light on the phenomenon of machine bias and are hopeful that it will ignite a debate about the need to augment current statistical translation tools with debiasing techniques -- which can already be found in the scientific literature.
\end{abstract}

\section{Introduction}

Although the idea of automated translation can in principle be traced back to as long as the 17th century with Ren\'{e} Descartes proposal of an ``universal language'' \cite{dascal1982universal}, machine translation has only existed as a technological field since the 1950s, with a pioneering memorandum by  Warren Weaver \cite{locke1955machine,weaver1955translation} discussing the possibility of employing digital computers to perform automated translation. The now famous Georgetown-IBM experiment followed not long after, providing the first experimental demonstration of the prospects of automating translation by the means of successfully converting more than sixty Russian sentences into English \cite{gordin2015scientific}. Early systems improved upon the results of the Georgetown-IBM experiment by exploiting Noam Chomsky's theory of generative linguistics, and the field experienced a sense of optimism about the prospects of fully automating natural language translation. As is customary with artificial intelligence, the initial optimistic stage was followed by an extended period of strong disillusionment with the field, of which the catalyst was the influential 1966 ALPAC (Automatic Language Processing Advisory Committee) report( \cite{hutchins1986machine}. 
Such research was then disfavoured in the United States, making a re-entrance in the 1970s before the 1980s surge in statistical methods for machine translation \cite{koehn2009statistical,Moses2007}. Statistical and example-based machine translation have been on the rise ever since \cite{Bahdanau2014,carl2003recent,Firat2017}, with highly successful applications such as Google Translate (recently ported to a neural translation technology \cite{wu2016google}) amounting to over $200$ million users daily.

In spite of the recent commercial success of automated translation tools (or perhaps stemming directly from it), machine translation has amounted a significant deal of criticism. Noted philosopher and founding father of generative linguistics Noam Chomsky has argued that the achievements of machine translation, while successes in a particular sense, are \emph{not successes in the sense that science has ever been interested in}: they merely provide effective ways, according to Chomsky, of approximating unanalyzed data \cite{Chomsky2011,norvig2017chomsky}. Chomsky argues that the faith of the MT community in statistical methods is absurd by analogy with a standard scientific field such as physics \cite{Chomsky2011}:

\begin{quotation}
\textsl{I mean actually you could do physics this way, instead of studying things like balls rolling down frictionless planes, which can't happen in nature, if you took a ton of video tapes of what's happening outside my office window, let's say, you know, leaves flying and various things, and you did an extensive analysis of them, you would get some kind of prediction of what's likely to happen next, certainly way better than anybody in the physics department could do. Well that's a notion of success which is I think novel, I don't know of anything like it in the history of science}.
\end{quotation}

Leading AI researcher and Google's Director of Research Peter Norvig responds to these arguments by suggesting that even standard physical theories such as the Newtonian model of gravitation are, in a sense, \emph{trained} \cite{norvig2017chomsky}:

\begin{quotation}
\textsl{As another example, consider the Newtonian model of gravitational attraction, which says that the force between two objects of mass $m_1$ and $m_2$ a distance $r$ apart is given by}
\begin{equation*}
F = G m_1 m_2 / r^2
\end{equation*}
\textsl{where $G$ is the universal gravitational constant. This is a trained model because the gravitational constant G is determined by statistical inference over the results of a series of experiments that contain stochastic experimental error. It is also a deterministic (non-probabilistic) model because it states an exact functional relationship. I believe that Chomsky has no objection to this kind of statistical model. Rather, he seems to reserve his criticism for statistical models like Shannon's that have quadrillions of parameters, not just one or two.}
\end{quotation}

Chomsky and Norvig's debate \cite{norvig2017chomsky} is a microcosm  of the two leading standpoints about the future of science in the face of increasingly sophisticated statistical models. Are we, as Chomsky seems to argue, jeopardizing science by relying on statistical tools to perform predictions instead of perfecting traditional science models, or are these tools, as Norvig argues, components of the scientific standard since its conception? Currently there are no satisfactory resolutions to this conundrum, but perhaps statistical models pose an even greater and more urgent threat to our society. 

On a 2014 article, Londa Schiebinger suggested that scientific research fails to take gender issues into account, arguing that the phenomenon of male defaults on new technologies such as Google Translate provides a window into this asymmetry \cite{schiebinger2014scientific}. Since then, recent worrisome results in machine learning have somewhat supported Schiebinger's view. Not only Google photos' statistical image labeling algorithm has been found to classify dark-skinned people as gorillas \cite{garcia2016racist} and purportedly intelligent programs have been suggested to be negatively biased against black prisoners when predicting criminal behavior \cite{angwin2016machine} but the machine learning revolution has also indirectly revived heated debates about the controversial field of physiognomy, with proposals of AI systems capable of identifying the sexual orientation of an individual through its facial characteristics \cite{wang2017deep}. Similar concerns are growing at an unprecedented rate in the media, with reports of Apple's Iphone X face unlock feature failing to differentiate between two different Asian people \cite{womanunlockphone2017} and automatic soap dispensers which reportedly do not recognize black hands \cite{racistsoapdispenser2017}. \emph{Machine bias}, the phenomenon by which trained statistical models unbeknownst to their creators grow to reflect controversial societal asymmetries, is growing into a pressing concern for the modern times, invites us to ask ourselves whether there are limits to our dependence on these techniques -- and more importantly, whether some of these limits have already been traversed. In the wave of algorithmic bias, some have argued for the creation of some kind of agency in the likes of the Food and Drug Administration, with the sole purpose of regulating algorithmic discrimination \cite{kirkpatrick2016battling}.

\begin{figure}[h]
	\centering
	\fbox{\includegraphics[width=\linewidth]{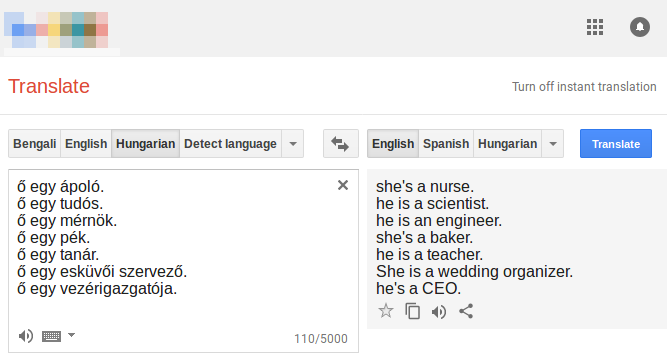}}
	\caption{Translating sentences from a gender neutral language such as Hungarian to English provides a glimpse into the phenomenon of gender bias in machine translation. This screenshot from Google Translate shows how occupations from traditionally male-dominated fields \cite{WB2014} such as scholar, engineer and CEO are interpreted as male, while occupations such as nurse, baker and wedding organizer are interpreted as female.}
	\label{fig:screenshot-gtranslate-hungarian}
\end{figure}

With this in mind, we propose a quantitative analysis of the phenomenon of gender bias in machine translation. We illustrate how this can be done by simply exploiting Google Translate to map sentences from a gender neutral language into English. As Figure \ref{fig:screenshot-gtranslate-hungarian} exemplifies, this approach produces results consistent with the hypothesis that sentences about stereotypical gender roles are translated accordingly with high probability: \emph{nurse} and \emph{baker} are translated with female pronouns while \emph{engineer} and \emph{CEO} are translated with male ones.

\section{Motivation}

As of 2018, Google Translate is one of the largest publicly available machine translation tools in existence, amounting 200 million users daily\cite{Gtranslate200daily2017}. Initially relying on United Nations and European Parliament transcripts to gather data, since 2014 Google Translate has inputed content from its users through the Translate Community initiative\cite{TranslateCommunity}. Recently however there has been a growing concern about gender asymmetries in the translation mechanism, with some heralding it as ``sexist'' \cite{AlgorithmGtranslateSexist2018}. This concern has to at least some extent a scientific backup: A recent study has shown that word embeddings are particularly prone to yielding gender stereotypes\cite{bolukbasi2016man}. Fortunately, the researchers propose a relatively simple \emph{debiasing} algorithm with promising results: they were able to cut the proportion of stereotypical analogies from $19\%$ to $6\%$ without any significant compromise in the performance of the word embedding technique. They are not alone: there is a growing effort to systematically discover and resolve issues of algorithmic bias in black-box algorithms\cite{hajian2016algorithmic}. The success of these results suggest that a similar technique could be used to remove gender bias from Google Translate outputs, should it exist. This paper intends to investigate whether it does. We are optimistic that our research endeavors can be used to argue that there is a positive payoff in redesigning modern statistical translation tools.

\section{Assumptions and Preliminaries}

In this paper we assume that a statistical translation tool should reflect at most the inequality existent in society -- it is only logical that a translation tool will poll from examples that  society produced and, as such, will inevitably retain some of that bias. It has been argued that one's language affects one's knowledge and cognition about the world \cite{kay1984sapir}, and this leads to the discussion that languages that distinguish between female and male genders grammatically may enforce a bias in the person's perception of the world, with some studies corroborating this, as shown in \cite{boroditsky2003sex}, as well some relating this with sexism \cite{thompson2014linguistic} and gender inequalities \cite{santacreu2013female}.

With this in mind, one can argue that a move towards gender neutrality in language and communication should be striven as a means to promote improved gender equality. Thus, in languages where gender neutrality can be achieved -- such as English -- it would be a valid aim to create translation tools that keep the gender-neutrality of texts translated into such a language, instead of defaulting to male or female variants.

We will thus assume throughout this paper that although the distribution of translated gender pronouns may deviate from 50:50, it should not deviate to the extent of misrepresenting the demographics of job positions. That is to say we shall assume that Google Translate incorporates a negative gender bias if the frequency of male defaults overestimates the (possibly unequal) distribution of male employees per female employee in a given occupation.

\section{Materials and Methods}

We shall assume and then show that the phenomenon of gender bias in machine translation can be assessed by mapping sentences constructed in gender neutral languages to English by the means of an automated translation tool. Specifically, we can translate sentences such as the Hungarian ``ő egy ápolónő'', where ``ápolónő'' translates to ``nurse'' and ``ő'' is a gender-neutral pronoun meaning either he, she or it, to English, yielding in this example the result ``she's a nurse'' on Google Translate. As Figure \ref{fig:screenshot-gtranslate-hungarian} clearly shows, the same template yields a male pronoun when ``nurse'' is replaced by ``engineer''. The same basic template can be ported to all other gender neutral languages, as depicted in Table \ref{tab:templates}. Given the success of Google Translate, which amounts to 200 million users daily, we have chosen to exploit its API to obtain the desired thermometer of gender bias. Also, in order to solidify our results, we have decided to work with a fair amount of gender neutral languages, forming a list of these with help from the World Atlas of Language Structures (WALS) \cite{wals} and other sources. Table \ref{tab:gender-neutral-languages} compiles all languages we chose to use, with additional columns informing whether they (1) exhibit a gender markers in the sentence and (2) are supported by Google Translate. However, we stumbled on some difficulties which led to some of those langauges being removed, which will be explained in . 

There is a prohibitively large class of nouns and adjectives that could in principle be substituted into our templates. To simplify our dataset, we have decided to focus our work on job positions -- which, we believe, are an interesting window into the nature of gender bias --, and were able to obtain a comprehensive list of professional occupations from the Bureau of Labor Statistics' detailed occupations table \cite{BLS2017}, from the United States Department of Labor. The values inside, however, had to be expanded since each line contained multiple occupations and sometimes very specific ones. Fortunately this table also provided a percentage of women participation in the jobs shown, for those that had more than 50 thousand workers. We filtered some of these because they were too generic ( ``Computer occupations, all other'', and others) or because they had gender specific words for the profession (``host/hostess'', ``waiter/waitress''). We then separated the curated jobs into broader categories (Artistic, Corporate, Theatre, etc.) as shown in Table \ref{tab:occupations}. Finally, Table \ref{tab:occupations-examples} shows thirty examples of randomly selected occupations from our dataset. For the occupations that had less than 50 thousand workers, and thus no data about the participation of women, we assumed that its women participation was that of its upper category. Finally, as complementary evidence we have decided to include a small subset of 21 adjectives in our study. All adjectives were obtained from the top one thousand most frequent words in this category as featured in the Corpus of Contemporary American English (COCA) https://corpus.byu.edu/coca/, but it was necessary to manually curate them because a substantial fraction of these adjectives cannot be applied to human subjects. Also because the sentiment associated with each adjective is not as easily accessible as for example the occupation category of each job position, we performed a manual selection of a subset of such words which we believe to be meaningful to this study. These words are presented in Table \ref{tab:adjectives}. We made all code and data used to generate and compile the results presented in the following sections publicly available in the following Github repository: https://github.com/marceloprates/Gender-Bias. Note however that because the Google Translate algorithm can change, unfortunately we cannot guarantee full reproducibility of our results. All experiments reported here were conducted on April 2018.

\begin{table}[H]
	\centering
	\begin{small}
		\begin{tabular}{|c|m{1.5cm}|m{1.5cm}|m{2.2cm}|m{1.5cm}|}
			\hline
			Language Family & Language & Phrases have male/female markers & Tested \\ \hline \hline
			Austronesian & Malay & \xmark & \checkmark\\ \hline
			\multirow{3}{*}{Uralic} & Estonian & \xmark & \checkmark\\
			& Finnish & \xmark & \checkmark\\
			& Hungarian & \xmark & \checkmark\\ \hline
			\multirow{5}{*}{Indo-European} & Armenian & \xmark & \checkmark\\
			& Bengali & O & \checkmark\\
			& English & \checkmark & \xmark \\
			& {\color{red}Persian} & \xmark & \checkmark\\
			& {\color{red}Nepali} & O  & \checkmark\\ \hline
			Japonic & Japanese & \xmark & \checkmark\\ \hline
			Koreanic & {\color{red}Korean} & \checkmark& \xmark\\ \hline
			Turkic & Turkish & \xmark & \checkmark\\ \hline
			\multirow{2}{*}{Niger-Congo} & Yoruba & \xmark & \checkmark\\
			& Swahili & \xmark & \checkmark\\ \hline
			Isolate & Basque & \xmark & \checkmark\\ \hline
			Sino-Tibetan & Chinese & O & \checkmark\\ \hline
		\end{tabular}
	\end{small}
	\caption{Gender neutral languages supported by Google Translate. Languages are grouped according to language families and classified according to whether they enforce any kind of mandatory gender (male/female) demarcation on simple phrases (\checkmark: yes, \xmark:~never, O:~some). For the purposes of this work, we have decided to work only with languages lacking such demarcation. Languages colored in red have been omitted for other reasons. See Section} \ref{ssec:rationale_for_language_exceptions} for further explanation.
	\label{tab:gender-neutral-languages}
\end{table}

\begin{table}[H]
\centering
\begin{small}
	\centering
	\begin{tabular}{|m{3.5cm}|m{2.0cm}|m{2.5cm}|m{2.5cm}|}
	\hline	
	Category 										& Group 						& \# Occupations 	& Female Participation	\\ \hline \hline
	Education, training, and library 				& Education						& $22$ 				& $73.0\%$				\\ \hline
	Business and financial operations				& Corporate						& $46$ 				& $54.0\%$				\\ \hline
	Office and administrative support				& Service						& $87$ 				& $72.2\%$				\\ \hline
	Healthcare support 								& Healthcare					& $16$ 				& $87.1\%$				\\ \hline
	Management 										& Corporate						& $46$ 				& $39.8\%$				\\ \hline
	Installation, maintenance, and repair 			& Service						& $91$ 				& $4.0\%$ 				\\ \hline
	Healthcare practitioners and technical 			& Healthcare					& $43$ 				& $75.0\%$				\\ \hline
	Community and social service 					& Service						& $14$ 				& $66.1\%$				\\ \hline
	Sales and related 								& Corporate						& $28$ 				& $49.1\%$				\\ \hline
	Production 										& Production					& $264$ 			& $28.9\%$				\\ \hline
	Architecture and engineering 					& STEM 							& $29$ 				& $16.2\%$				\\ \hline
	Life, physical, and social science				& STEM 							& $34$ 				& $47.4\%$				\\ \hline
	Transportation and material moving				& Service						& $70$ 				& $17.3\%$				\\ \hline
	Arts, design, entertainment, sports, and media 	& Arts / Entertainment 			& $37$ 				& $46.9\%$				\\ \hline
	Legal											& Legal 						& $7$				& $52.8\%$				\\ \hline
	Protective Service 								& Service 						& $28$ 				& $22.3\%$				\\ \hline
	Food preparation and serving related 			& Service 						& $17$ 				& $53.8\%$				\\ \hline
	Farming, fishing, and forestry 					& Farming / Fishing / Forestry 	& $13$ 				& $23.4\%$				\\ \hline
	Computer and mathematical						& STEM 							& $16$ 				& $25.5\%$				\\ \hline
	Personal care and service						& Service 						& $33$ 				& $76.1\%$				\\ \hline
	Construction and extraction 					& Construction / Extraction 	& $68$ 				& $3.0\%$ 				\\ \hline
	Building and grounds cleaning and maintenance 	& Service 						& $10$ 				& $40.7\%$				\\ \hline \hline
	Total											& - 							& $1019$			& $41.3\%$				\\ \hline
	\end{tabular}
    \end{small}
	\caption{Selected occupations obtained from the U.S. Bureau of Labor Statistics https://www.bls.gov/cps/cpsaat11.htm, grouped by category. We obtained a total of $1019$ occupations from $22$ distinct categories. We have further grouped them into broader groups (or \emph{super-categories}) to ease analysis and visualization.}
	\label{tab:occupations}
\end{table}

\begin{table}[H]
	\centering
	\small{
	\begin{tabular}{|c|c|c|}
		\hline
		Language & Occupation sentence template & Adjective sentence template \\ \hline \hline
		Malay & dia adalah $\langle occupation \rangle$ & dia $\langle adjective \rangle$ \\ \hline
		Estonian & ta on $\langle occupation \rangle$ & ta on $\langle adjective \rangle$ \\ \hline
		Finnish & hän on $\langle occupation \rangle$ & hän on $\langle adjective \rangle$ \\ \hline
		Hungarian & ő egy $\langle occupation \rangle$ & ő $\langle adjective \rangle$ \\ \hline
		Armenian & na $\langle occupation \rangle$ e & na $\langle adjective \rangle$ e \\ \hline
		\multirow{6}{*}{Bengali} & Ē ēkajana $\langle occupation \rangle$ & Ē $\langle adjective \rangle$\\
		 & Yini ēkajana $\langle occupation \rangle$ & Yini $\langle adjective \rangle$\\
		 & Ō ēkajana $\langle occupation \rangle$ & Ō $\langle adjective \rangle$\\
		 & Uni ēkajana $\langle occupation \rangle$ & Uni $\langle adjective \rangle$\\
		 & Sē ēkajana $\langle occupation \rangle$ & Sē $\langle adjective \rangle$\\
		 & Tini ēkajana $\langle occupation \rangle$ & Tini $\langle adjective \rangle$\\ \hline
		Japanese & \begin{CJK*}{UTF8}{min}あの人は\end{CJK*} $\langle occupation \rangle$ \begin{CJK*}{UTF8}{min}です\end{CJK*} & \begin{CJK*}{UTF8}{min}あの人は\end{CJK*} $\langle adjective \rangle$ \begin{CJK*}{UTF8}{min}です\end{CJK*}\\ \hline
		Turkish & o bir $\langle occupation \rangle$ & o $\langle adjective \rangle$ \\ \hline
		Yoruba & o jẹ $\langle occupation \rangle$ & o jẹ $\langle adjective \rangle$ \\ \hline
		Basque & $\langle occupation \rangle$ bat da & $\langle adjective \rangle$ da \\ \hline
		Swahili & yeye ni $\langle occupation \rangle$ & yeye ni $\langle adjective \rangle$ \\ \hline
		Chinese & ta shi $\langle occupation \rangle$ & ta hen $\langle adjective \rangle$ \\ \hline
	\end{tabular}
	}
	\caption{Templates used to infer gender biases in the translation of job occupations and adjectives to the English language.}
	\label{tab:templates}
\end{table}

\begin{table}[H]
\centering
\small{
	\begin{tabular}{|c|c|c|c|c|}
	\hline
	Insurance sales agent 	& Editor 						& Rancher 					\\ \hline
	Ticket taker 			& Pile-driver operator 			& Tool maker 				\\ \hline
	Jeweler 				& Judicial law clerk 			& Auditing clerk 			\\ \hline
	Physician 				& Embalmer 						& Door-to-door salesperson 	\\ \hline
	Packer 					& Bookkeeping clerk 			& Community health worker 	\\ \hline
	Sales worker 			& Floor finisher 				& Social science technician \\ \hline
	Probation officer 		& Paper goods machine setter 	& Heating installer 		\\ \hline
	Animal breeder 			& Instructor 					& Teacher assistant 		\\ \hline
	Statistical assistant 	& Shipping clerk 				& Trapper 					\\ \hline
	Pharmacy aide 			& Sewing machine operator 		& Service unit operator 	\\ \hline
	\end{tabular}
	}
    \caption{A randomly selected example subset of thirty occupations obtained from our dataset with a total of $1019$ different occupations.}
	\label{tab:occupations-examples}
\end{table}

\begin{table}[H]
\centering
\small{
	\begin{tabular}{|c|c|c|}
	\hline
	Happy 		& Sad 		 & Right 		\\ \hline
	Wrong 		& Afraid	 & Brave 		\\ \hline
	Smart		& Dumb		 & Proud 		\\ \hline
	Strong		& Polite	 & Cruel 		\\ \hline
	Desirable	& Loving	 & Sympathetic 	\\ \hline
	Modest 		& Successful & Guilty		\\ \hline
	Innocent	& Mature	 & Shy			\\ \hline
	\end{tabular}
	}
    \caption{Curated list of 21 adjectives obtained from the top one thousand most frequent words in this category in the Corpus of Contemporary American English (COCA)} https://corpus.byu.edu/coca/.
	\label{tab:adjectives}
\end{table}

\subsection{Rationale for language exceptions}
\label{ssec:rationale_for_language_exceptions}

While it is possible to construct gender neutral sentences in two of the languages omitted in our experiments (namely Korean and Nepali), we have chosen to omit them for the following reasons:

\begin{enumerate}
\item We faced technical difficulties to form templates and automatically translate sentences with the right-to-left, top-to-bottom nature of the script and, as such, we have decided not to include it in our experiments.
\item Due to Nepali having a rather complex grammar, with possible male/female gender demarcations on the phrases and due to none of the authors being fluent or able to reach someone fluent in the language, we were not confident enough in our ability to produce the required templates. Bengali was almost discarded under the same rationale, but we have decided to keep it because of our sentence template for Bengali has a simple grammatical structure which does not require any kind of inflection.
\item One can construct gender neutral phrases in Korean by omitting the gender pronoun; in fact, this is the default procedure. However, the expressiveness of this omission depends on the context of the sentence being clear, which is not possible in the way we frame phrases.
\end{enumerate}

\section{Distribution of translated gender pronouns per occupation category}

A sensible way to group translation data is to coalesce occupations in the same category and collect statistics among languages about how prominent male defaults are in each field. What we have found is that Google Translate does indeed translate sentences with male pronouns with greater probability than it does either with female or gender-neutral pronouns, in general. Furthermore, this bias is seemingly aggravated for fields suggested to be troubled by male stereotypes, such as life and physical sciences, architecture, engineering, computer science and mathematics \cite{moss2015can}. Table \ref{tab:gender-by-category} summarizes these data, and Table \ref{tab:gender-by-category-grouped} summarizes it even further by coalescing occupation categories into broader groups to ease interpretation. For instance, STEM (Science, Technology, Engineering and Mathematics) fields are grouped into a single category, which helps us compare the large asymmetry between gender pronouns in these fields ($72\%$ of male defaults) to that of more evenly distributed fields such as healthcare ($50\%$).

\begin{table}[H]
\footnotesize\setlength{\tabcolsep}{2.5pt}
\centering
\small{
	\begin{tabular}{|c|c|c|c|}
	\hline
	Category & Female ($\%$)	& Male ($\%$)	& Neutral ($\%$) \\ \hline
	\hline
	Office and administrative support & 11.015 & 58.812 & 16.954 \\ \hline
	Architecture and engineering & 2.299 & 72.701 & 10.92 \\ \hline
	Farming, fishing, and forestry & 12.179 & 62.179 & 14.744 \\ \hline
	Management & 11.232 & 66.667 & 12.681 \\ \hline
	Community and social service & 20.238 & 62.5 & 10.119 \\ \hline
	Healthcare support & 25.0 & 43.75 & 17.188 \\ \hline
	Sales and related & 8.929 & 62.202 & 16.964 \\ \hline
	Installation, maintenance, and repair & 5.22 & 58.333 & 17.125 \\ \hline
	Transportation and material moving & 8.81 & 62.976 & 17.5 \\ \hline
	Legal & 11.905 & 72.619 & 10.714 \\ \hline
	Business and financial operations & 7.065 & 67.935 & 15.58 \\ \hline
	Life, physical, and social science & 5.882 & 73.284 & 10.049 \\ \hline
	Arts, design, entertainment, sports, and media & 10.36 & 67.342 & 11.486 \\ \hline
	Education, training, and library & 23.485 & 53.03 & 9.091 \\ \hline
	Building and grounds cleaning and maintenance & 12.5 & 68.333 & 11.667 \\ \hline
	Personal care and service & 18.939 & 49.747 & 18.434 \\ \hline
	Healthcare practitioners and technical & 22.674 & 51.744 & 15.116 \\ \hline
	Production & 14.331 & 51.199 & 18.245 \\ \hline
	Computer and mathematical & 4.167 & 66.146 & 14.062 \\ \hline
	Construction and extraction & 8.578 & 61.887 & 17.525 \\ \hline
	Protective service & 8.631 & 65.179 & 12.5 \\ \hline
	Food preparation and serving related & 21.078 & 58.333 & 17.647 \\ \hline \hline
	Total & 11.76 & 58.93 & 15.939 \\ \hline
	\end{tabular}
	}
    \caption{Percentage of female, male and neutral gender pronouns obtained for each BLS occupation category, averaged over all occupations in said category and tested languages detailed in Table} \ref{tab:gender-neutral-languages}. Note that rows do not in general add up to $100\%$, as there is a fair amount of translated sentences for which we cannot obtain a gender pronoun.
	\label{tab:gender-by-category}
\end{table}

\begin{table}[H]
\centering
\small{
	\begin{tabular}{|c|c|c|c|}
	\hline
	Category 								& Female ($\%$)			& Male ($\%$)			& Neutral ($\%$)	\\ \hline
	\hline
	Service & 10.5 & 59.548 & 16.476 \\ \hline
	STEM & 4.219 & 71.624 & 11.181 \\ \hline
	Farming / Fishing / Forestry & 12.179 & 62.179 & 14.744 \\ \hline
	Corporate & 9.167 & 66.042 & 14.861 \\ \hline
	Healthcare & 23.305 & 49.576 & 15.537 \\ \hline
	Legal & 11.905 & 72.619 & 10.714 \\ \hline
	Arts / Entertainment & 10.36 & 67.342 & 11.486 \\ \hline
	Education & 23.485 & 53.03 & 9.091 \\ \hline
	Production & 14.331 & 51.199 & 18.245 \\ \hline
	Construction / Extraction & 8.578 & 61.887 & 17.525 \\ \hline \hline
	Total & 11.76 & 58.93 & 15.939 \\ \hline

	\end{tabular}
	}
    \caption{Percentage of female, male and neutral gender pronouns obtained for each of the merged occupation category, averaged over all occupations in said category and tested languages detailed in Table} \ref{tab:gender-neutral-languages}. Note that rows do not in general add up to $100\%$, as there is a fair amount of translated sentences for which we cannot obtain a gender pronoun.
	\label{tab:gender-by-category-grouped}
\end{table}

\definecolor{beige}{HTML}{D3CA39}

\begin{figure}[H]
	\centering
	\includegraphics[width=10cm]{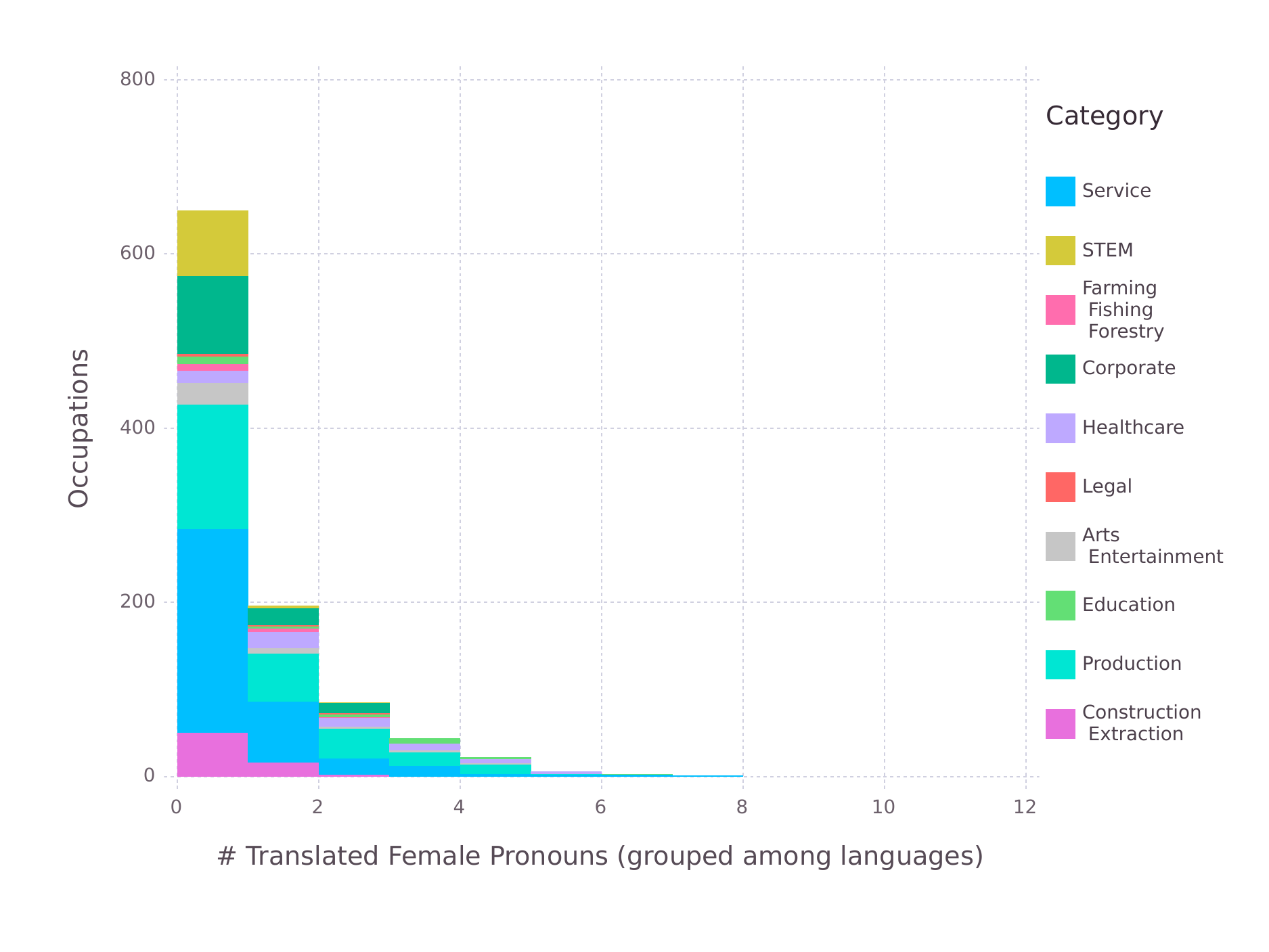}
	\caption{The data for the number of translated female pronouns per merged occupation category totaled among languages suggests and inverse distribution. STEM fields are nearly exclusively concentrated at $X = 0$, while more evenly distributed in fields such as production and healthcare (See Table}~\ref{tab:gender-by-category-grouped}) extends to higher values.
	\label{fig:histogram-female-grouped}
\end{figure}

\begin{figure}[H]
	\centering
	\includegraphics[width=10cm]{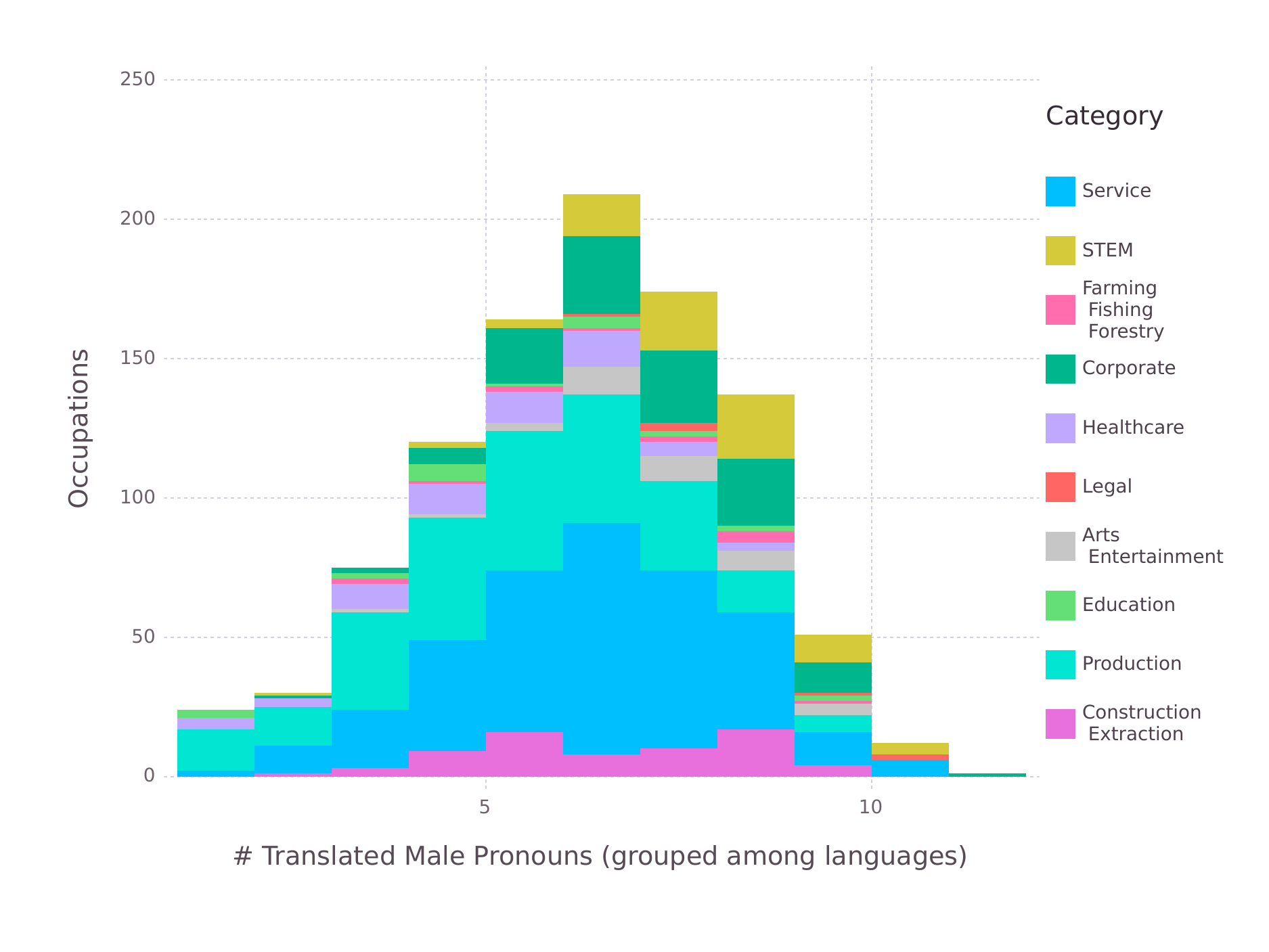}
	\caption{In contrast to Figure}\ref{fig:histogram-female-grouped} male pronouns are seemingly skew normally distributed, with a peak at $X = 6$. One can see how STEM fields concentrate mainly to the right ($X \geq 6$).
	\label{fig:histogram-male-grouped}
\end{figure}

\begin{figure}[H]
	\centering
	\includegraphics[width=10cm]{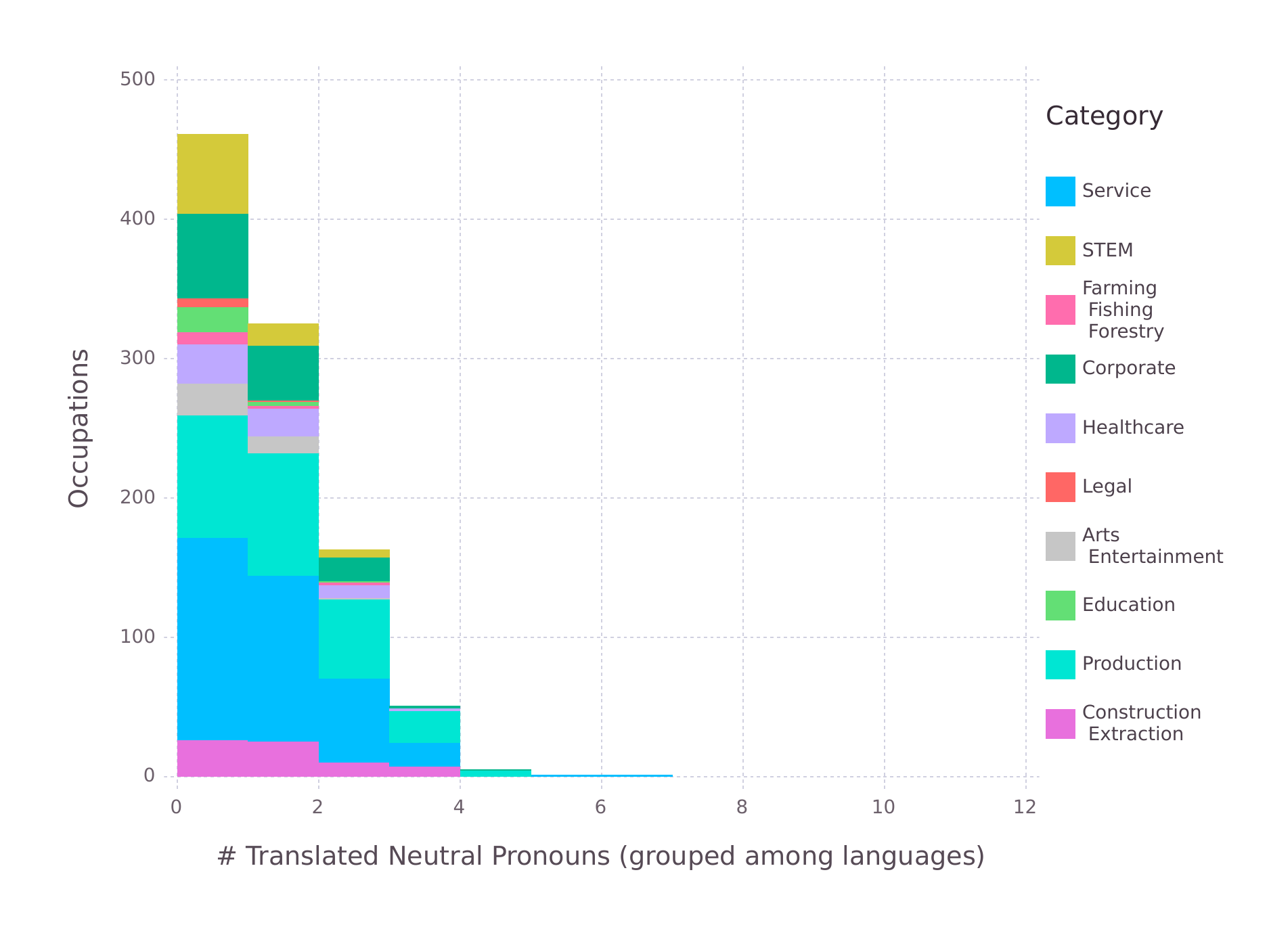}
	\caption{The scarcity of gender-neutral pronouns is manifest in their histogram. Once again, STEM fields are predominantly concentrated at $X = 0$.}
	\label{fig:histogram-neutral-grouped}
\end{figure}

Plotting histograms for the number of gender pronouns per occupation category sheds further light on how female, male and gender-neutral pronouns are differently distributed. The histogram in Figure~\ref{fig:histogram-female-grouped} suggests that the number of female pronouns is inversely distributed -- which is mirrored in the data for gender-neutral pronouns in Figure~\ref{fig:histogram-neutral-grouped} --, while the same data for male pronouns (shown in Figure~\ref{fig:histogram-male-grouped}) suggests a skew normal distribution. Furthermore we can see both on Figures \ref{fig:histogram-female-grouped} and \ref{fig:histogram-male-grouped} how STEM fields (labeled in \colorbox{beige}{beige}
exhibit predominantly male defaults -- amounting predominantly near $X = 0$ in the female histogram although much to the right in the male histogram.

These values contrast with BLS' report of gender participation, which will be discussed in more detail in Section \ref{sec:comparison-women-participation}.

\begin{figure}[H]
	\centering
	\includegraphics[width=10cm]{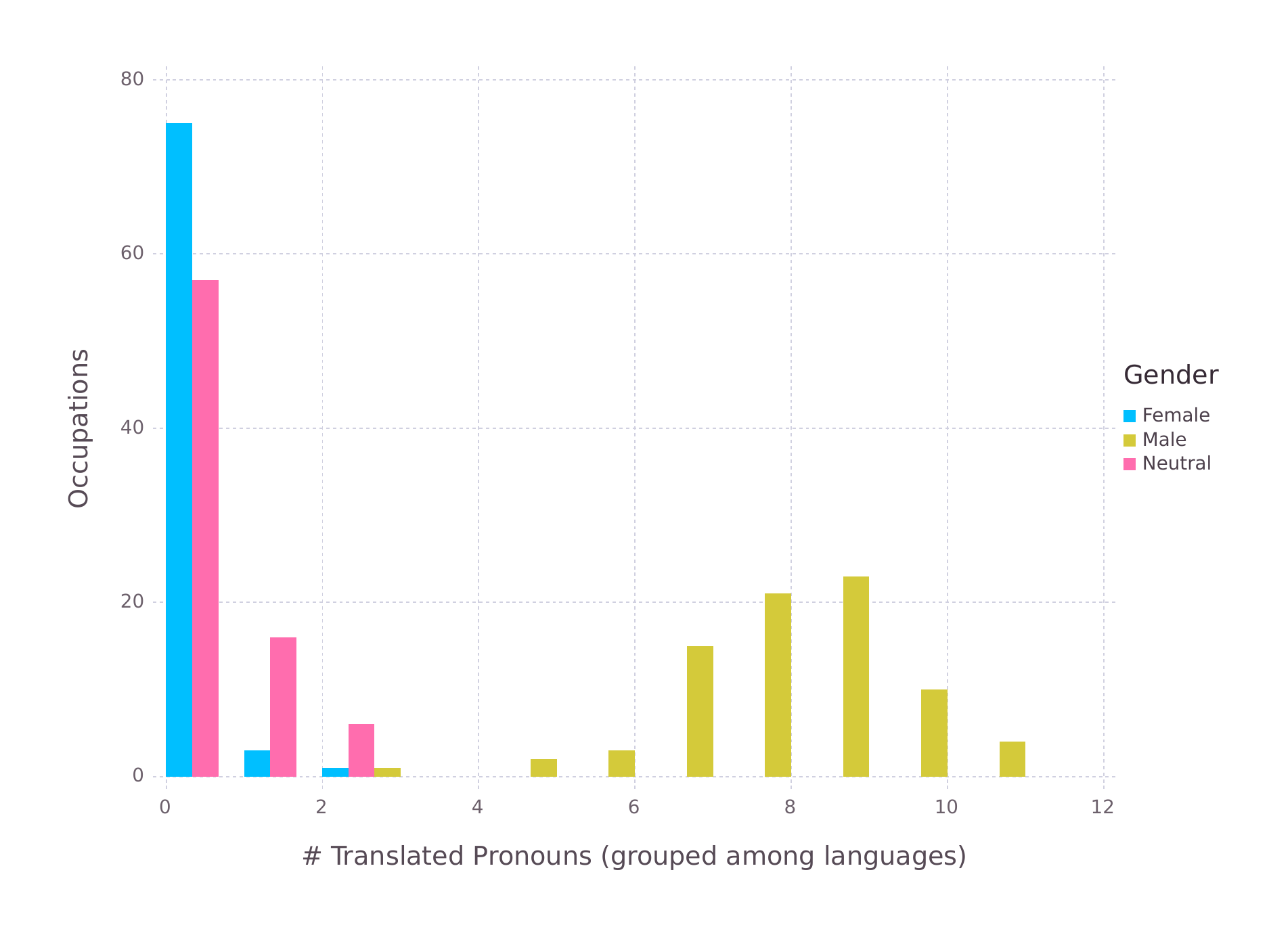}
	\caption{Histograms for the distribution of the number of translated female, male and gender neutral pronouns totaled among languages are plotted side by side for job occupations in the STEM (Science, Technology, Engineering and Mathematics) field, in which male defaults are the second-to-most prominent (after Legal).}
	\label{fig:histogram-dodged-STEM}
\end{figure}

\begin{figure}[H]
	\centering
	\includegraphics[width=10cm]{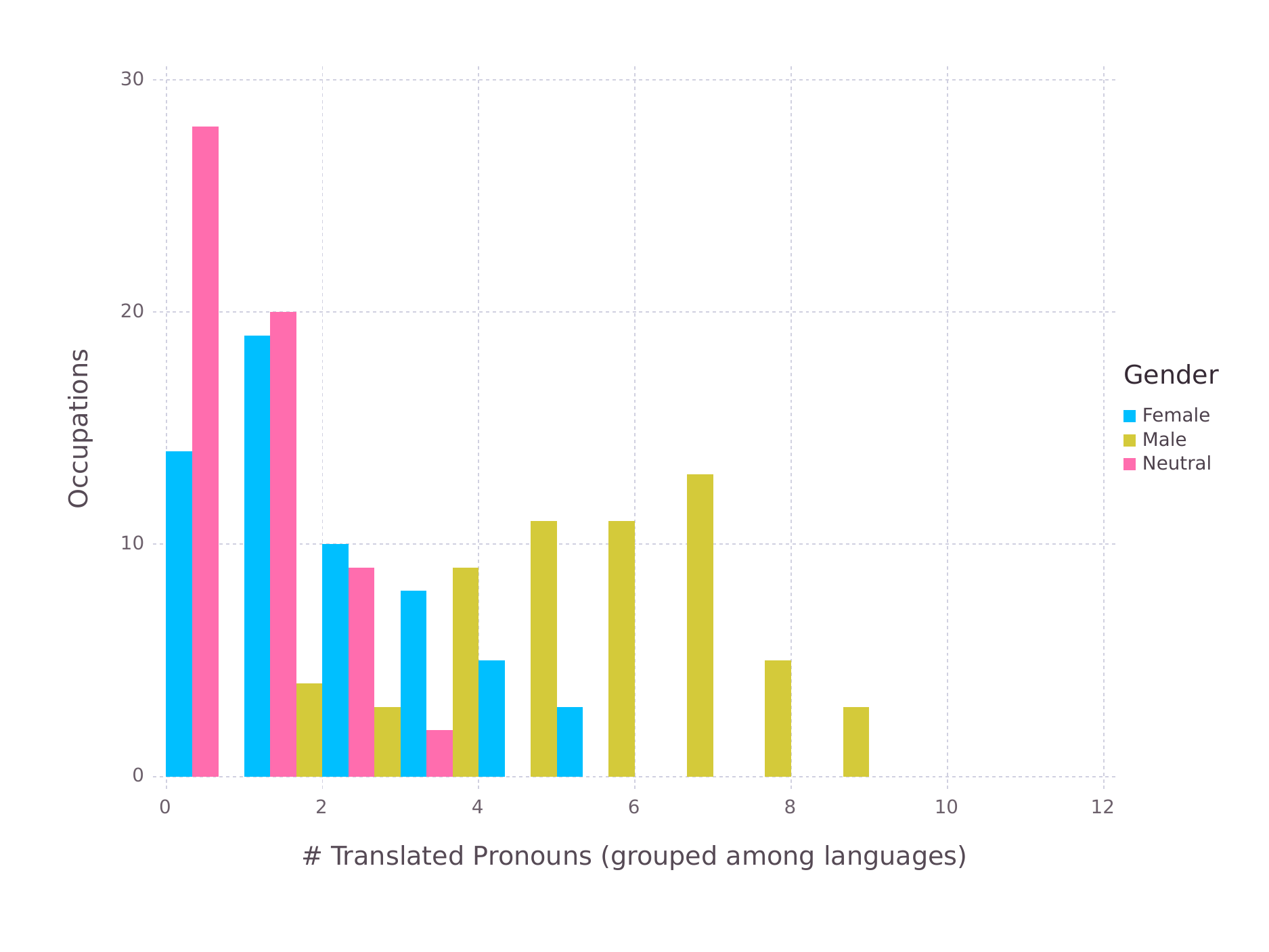}
	\caption{Histograms for the distribution of the number of translated female, male and gender neutral pronouns totaled among languages are plotted side by side for job occupations in the Healthcare field, in which male defaults are least prominent.}
	\label{fig:histogram-dodged-Healthcare}
\end{figure}

We can also visualize male, female, and gender neutral histograms side by side, in which context is useful to compare the dissimilar distributions of translated STEM and Healthcare occupations (Figures \ref{fig:histogram-dodged-STEM} and \ref{fig:histogram-dodged-Healthcare} respectively). The number of translated female pronouns among languages is not normally distributed for any of the individual categories in Table \ref{tab:occupations}, but Healthcare is in many ways the most balanced category, which can be seen in comparison with STEM -- in which male defaults are second to most prominent.

\begin{figure}[H]
	\centering
	\includegraphics[width=10cm]{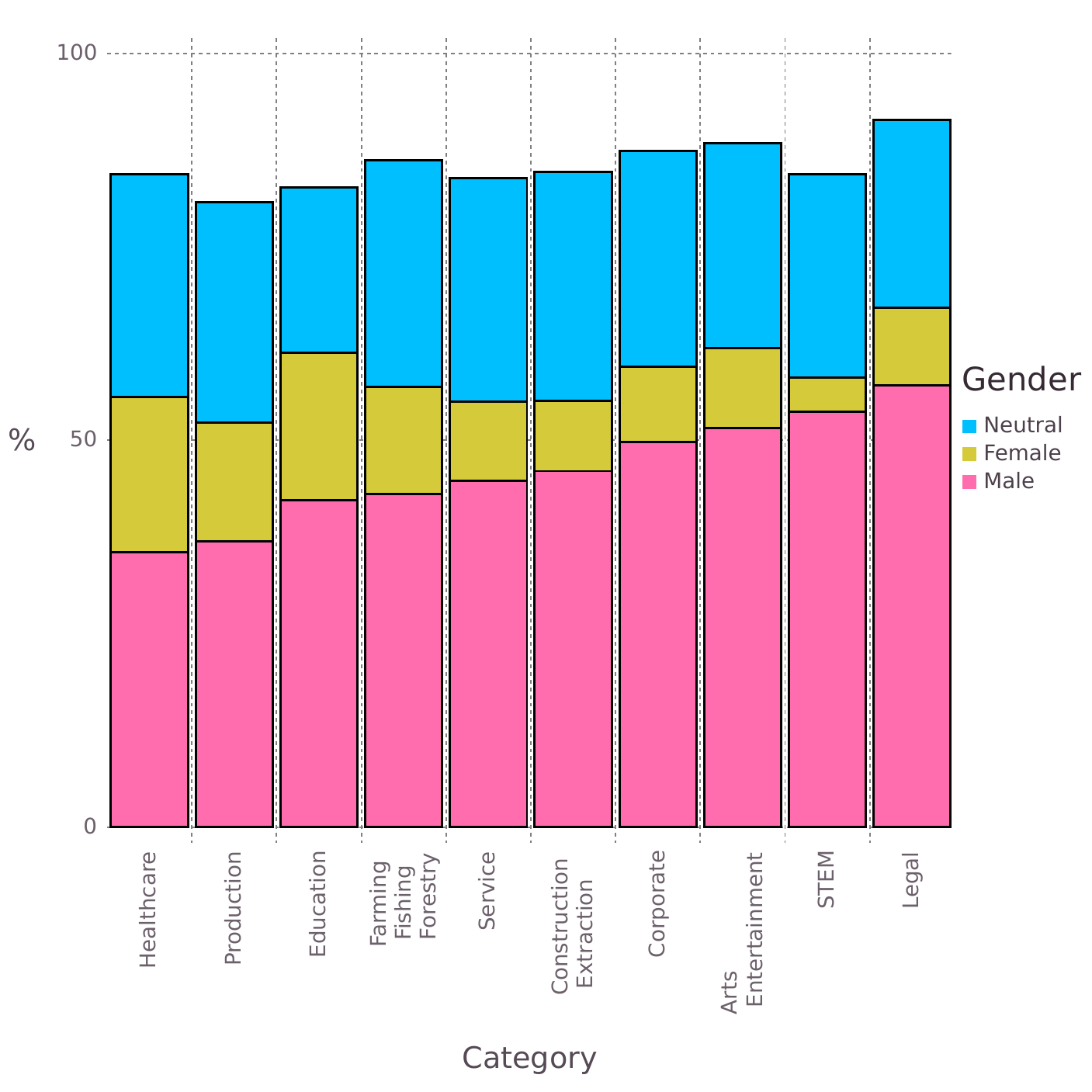}
	\caption{Bar plots show how much of the distribution of translated gender pronouns for each occupation category (grouped as in Table \ref{tab:gender-by-category-grouped}) is composed of female, male and neutral terms. Legal and STEM fields exhibit a predominance of male defaults and contrast with Healthcare and Education, with a larger proportion of female and neutral pronouns. Note that in general the bars do not add up to $100\%$, as there is a fair amount of translated sentences for which we cannot obtain a gender pronoun. Categories are sorted with respect to the proportions of male, female and neutral translated pronouns respectively}.
	\label{fig:gender-by-category}
\end{figure}

The bar plots in Figure \ref{fig:gender-by-category} help us visualize how much of the distribution of each occupation category is composed of female, male and gender-neutral pronouns. In this context, STEM fields, which show a predominance of male defaults, are contrasted with Healthcare and educations, which show a larger proportion of female pronouns.

\begin{figure}[H]
	\centering
	\includegraphics[width=10cm]{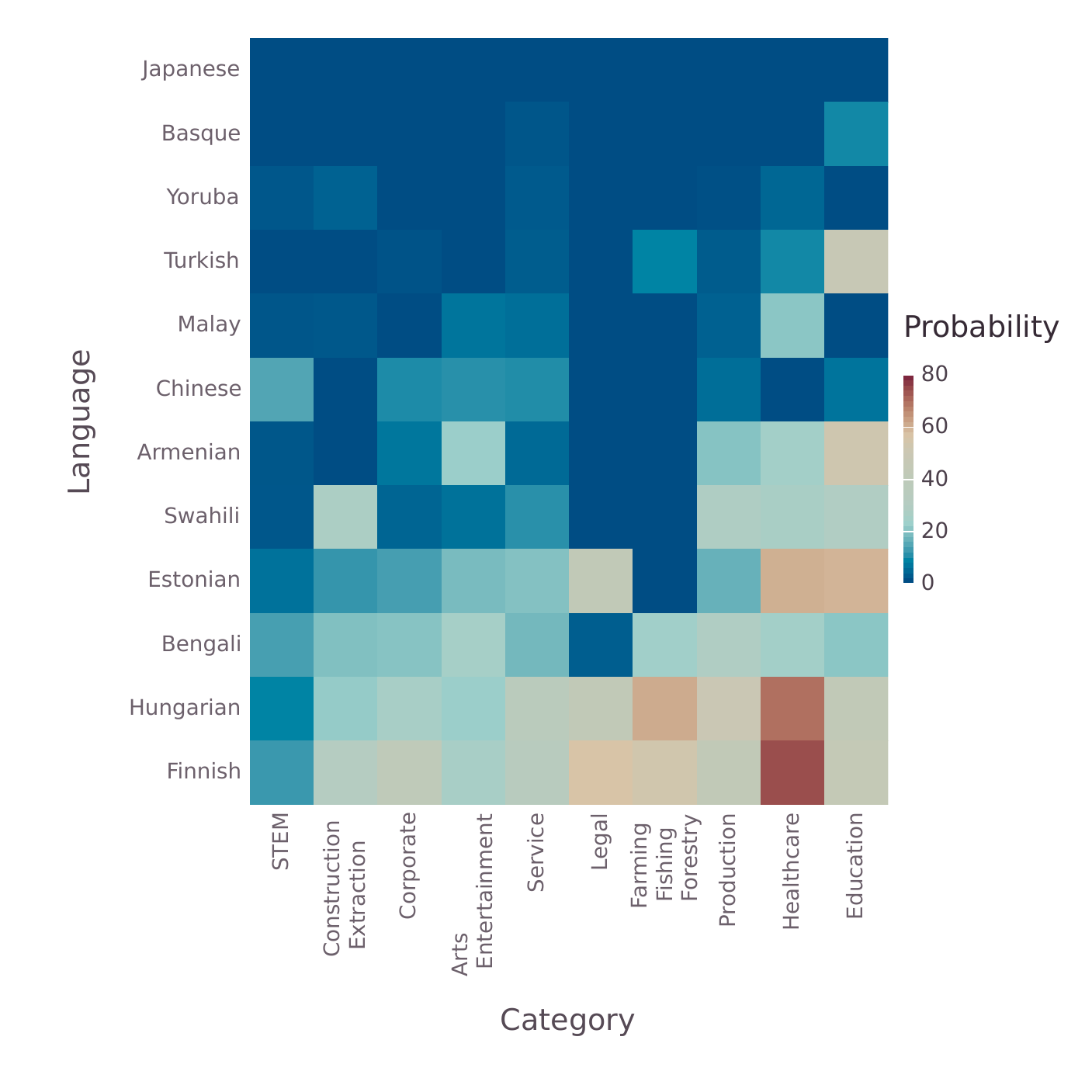}
	\caption{Heatmap for the translation probability into female pronouns for each pair of language and occupation category. Probabilities range from $0\%$ (blue) to $100\%$ (red), and both axes are sorted in such a way that higher probabilities concentrate on the bottom right corner.}
	\label{fig:heatmap-female}
\end{figure}

\begin{figure}[H]
	\centering
	\includegraphics[width=10cm]{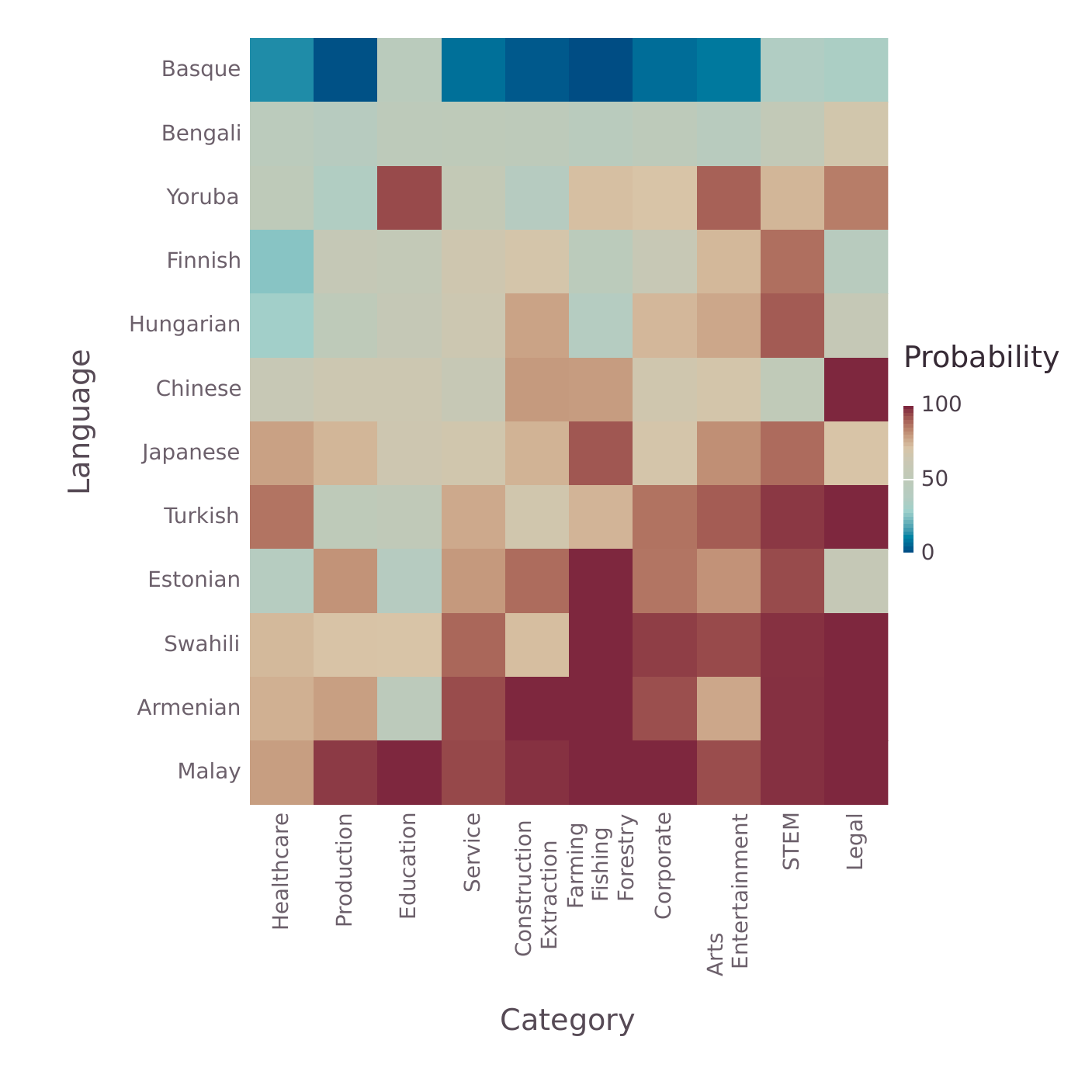}
	\caption{Heatmap for the translation probability into male pronouns for each pair of language and occupation category. Probabilities range from $0\%$ (blue) to $100\%$ (red), and both axes are sorted in such a way that higher probabilities concentrate on the bottom right corner.}
	\label{fig:heatmap-male}
\end{figure}

\begin{figure}[H]
	\centering
	\includegraphics[width=10cm]{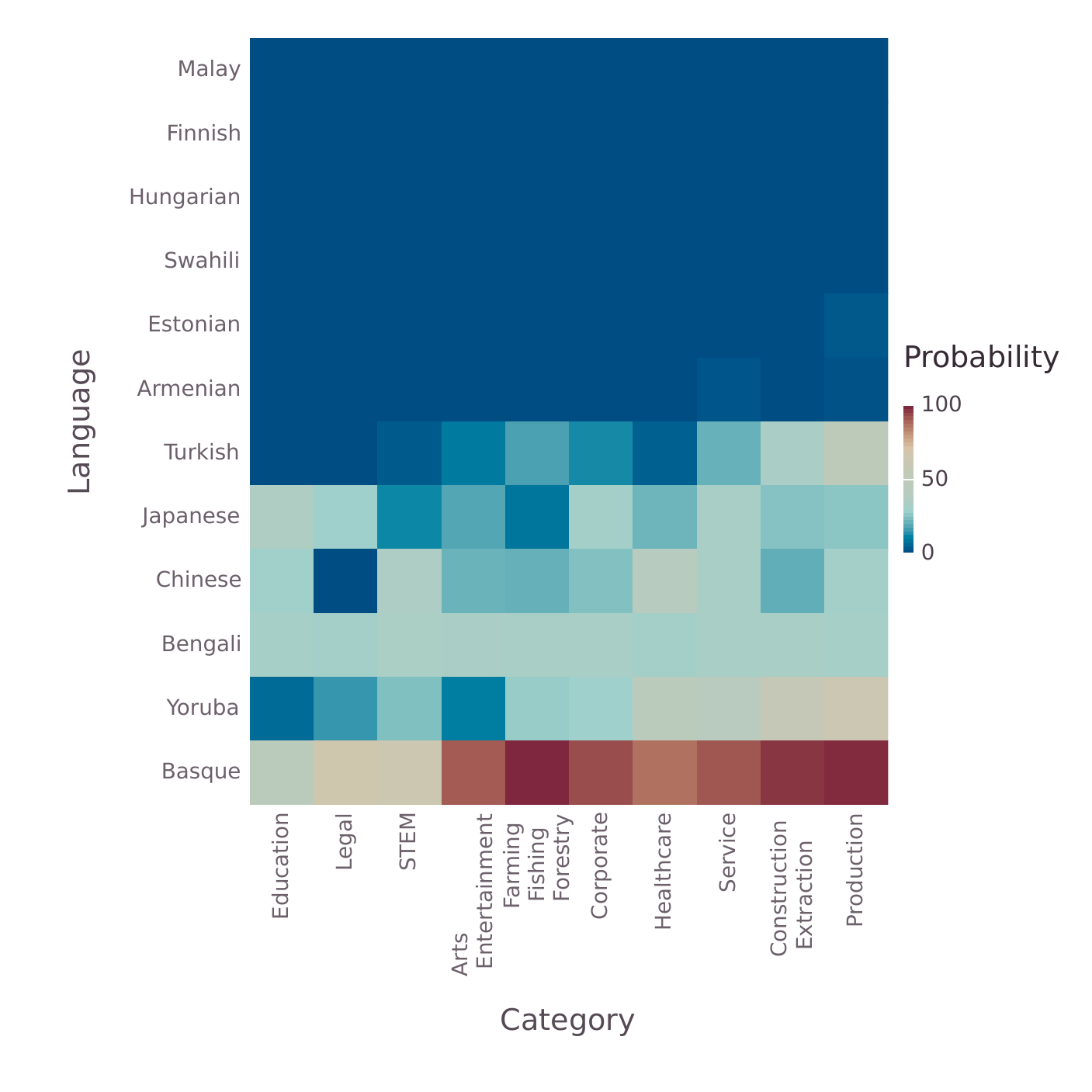}
	\caption{Heatmap for the translation probability into gender neutral pronouns for each pair of language and occupation category. Probabilities range from $0\%$ (blue) to $100\%$ (red), and both axes are sorted in such a way that higher probabilities concentrate on the bottom right corner.}
	\label{fig:heatmap-neutral}
\end{figure}

Although computing our statistics over the set of all languages has practical value, this may erase subtleties characteristic to each individual idiom. In this context, it is also important to visualize how each language translates job occupations in each category. The heatmaps in Figures  \ref{fig:heatmap-female}, \ref{fig:heatmap-male} and \ref{fig:heatmap-neutral} show the translation probabilities into female, male and neutral pronouns, respectively, for each pair of language and category (blue is $0\%$ and red is $100\%$). Both axes are sorted in these Figures, which helps us visualize both languages and categories in an spectrum of increasing male/female/neutral translation tendencies. In agreement with suggested stereotypes, \cite{moss2015can} STEM fields are second only to Legal ones in the prominence of male defaults. These two are followed by Arts \& Entertainment and Corporate, in this order, while Healthcare, Production and Education lie on the opposite end of the spectrum.

Our analysis is not truly complete without tests for statistical significant differences in the translation tendencies among female, male and gender neutral pronouns. We want to know for which languages and categories does Google Translate translate sentences with significantly more male than female, or male than neutral, or neutral than female, pronouns. We ran one-sided t-tests to assess this question for each pair of language and category and also totaled among either languages or categories. The corresponding p-values are presented in Tables \ref{tab:pvalues-MF}, \ref{tab:pvalues-MN}, \ref{tab:pvalues-NF} respectively. Language-Category pairs for which the null hypothesis was not rejected for a confidence level of $\alpha = .005$ are highlighted in blue. It is important to note that when the null hypothesis is accepted, we cannot discard the possibility of the complementary null hypothesis being rejected. For example, neither male nor female pronouns are significantly more common for Healthcare positions in the Estonian language, but female pronouns \emph{are} significantly more common for the same category in Finnish and Hungarian. Because of this, Language-Category pairs for which the complementary null hypothesis is rejected are painted in a darker shade of blue (see Table \ref{tab:pvalues-MF} for the three examples cited above.

Although there is a noticeable level of variation among languages and categories, the null hypothesis that male pronouns are not significantly more frequent than female ones was consistently rejected for all languages and all categories examined. The same is true for the null hypothesis that male pronouns are not significantly more frequent than gender neutral pronouns, with the one exception of the Basque language (which exhibits a rather strong tendency towards neutral pronouns). The null hypothesis that neutral pronouns are not significantly more frequent than female ones is accepted with much more frequency, namely for the languages Malay, Estonian, Finnish, Hungarian, Armenian and for the categories Farming \& Fishing \& Forestry, Healthcare, Legal, Arts \& Entertainment, Education. In all three cases, the null hypothesis corresponding to the aggregate for all languages and categories is rejected. We can learn from this, in summary, that Google Translate translates male pronouns more frequently than both female and gender neutral ones, either in general for Language-Category pairs or consistently among languages and among categories (with the notable exception of the Basque idiom).

\begin{table}[H]
\centering
\footnotesize\setlength{\tabcolsep}{2.5pt}
\small
\begin{tabular}{|m{1.75cm}|cccccccccccc|c|}
\hline
& Mal. & Est. & Fin. & Hun. & Arm. & Ben. & Jap. & Tur. & Yor. & Bas. & Swa. & Chi. & Total \\ \hline
Service	&	$<\alpha$	&	$<\alpha$	&	$<\alpha$	&	$<\alpha$	&	$<\alpha$	&	$<\alpha$	&	$<\alpha$	&	$<\alpha$	&	$<\alpha$	&	$<\alpha$	&	$<\alpha$	&	$<\alpha$	&	$<\alpha$	 \\ \hline 
STEM	&	$<\alpha$	&	$<\alpha$	&	$<\alpha$	&	$<\alpha$	&	$<\alpha$	&	$<\alpha$	&	$<\alpha$	&	$<\alpha$	&	$<\alpha$	&	$<\alpha$	&	$<\alpha$	&	$<\alpha$	&	$<\alpha$	 \\ \hline 
Farming 
 Fishing 
 Forestry	&	$<\alpha$	&	$<\alpha$	&	\cellcolor{blue!25}$.603$	&	\cellcolor{blue!25}$.786$	&	$<\alpha$	&	$<\alpha$	&	$<\alpha$	&	$<\alpha$	&	$<\alpha$	&	\cellcolor{blue!25}$*$	&	$<\alpha$	&	$<\alpha$	&	$<\alpha$	 \\ \hline 
Corporate	&	$<\alpha$	&	$<\alpha$	&	$<\alpha$	&	$<\alpha$	&	$<\alpha$	&	$<\alpha$	&	$<\alpha$	&	$<\alpha$	&	$<\alpha$	&	$<\alpha$	&	$<\alpha$	&	$<\alpha$	&	$<\alpha$	 \\ \hline 
Healthcare	&	$<\alpha$	&	\cellcolor{blue!25}$.938$	&	\cellcolor{blue!45}$1.0$	&	\cellcolor{blue!45}$.999$	&	$<\alpha$	&	$<\alpha$	&	$<\alpha$	&	$<\alpha$	&	$<\alpha$	&	$<\alpha$	&	$<\alpha$	&	$<\alpha$	&	$<\alpha$	 \\ \hline 
Legal	&	$<\alpha$	&	\cellcolor{blue!25}$.368$	&	\cellcolor{blue!25}$.632$	&	\cellcolor{blue!25}$.368$	&	$<\alpha$	&	$<\alpha$	&	$<\alpha$	&	$<\alpha$	&	$<\alpha$	&	\cellcolor{blue!25}$.086$	&	$<\alpha$	&	$<\alpha$	&	$<\alpha$	 \\ \hline 
Arts 
 Entertainment	&	$<\alpha$	&	$<\alpha$	&	$<\alpha$	&	$<\alpha$	&	$<\alpha$	&	$<\alpha$	&	$<\alpha$	&	$<\alpha$	&	$<\alpha$	&	\cellcolor{blue!25}$.08$	&	$<\alpha$	&	$<\alpha$	&	$<\alpha$	 \\ \hline 
Education	&	$<\alpha$	&	\cellcolor{blue!25}$.808$	&	\cellcolor{blue!25}$.333$	&	\cellcolor{blue!25}$.263$	&	\cellcolor{blue!25}$.588$	&	$<\alpha$	&	$<\alpha$	&	\cellcolor{blue!25}$.417$	&	$<\alpha$	&	\cellcolor{blue!25}$.052$	&	$<\alpha$	&	$<\alpha$	&	$<\alpha$	 \\ \hline 
Production	&	$<\alpha$	&	$<\alpha$	&	$<\alpha$	&	\cellcolor{blue!25}$.5$	&	$<\alpha$	&	$<\alpha$	&	$<\alpha$	&	$<\alpha$	&	$<\alpha$	&	\cellcolor{blue!25}$.159$	&	$<\alpha$	&	$<\alpha$	&	$<\alpha$	 \\ \hline 
Construction 
 Extraction	&	$<\alpha$	&	$<\alpha$	&	$<\alpha$	&	$<\alpha$	&	$<\alpha$	&	$<\alpha$	&	$<\alpha$	&	$<\alpha$	&	$<\alpha$	&	\cellcolor{blue!25}$.16$	&	$<\alpha$	&	$<\alpha$	&	$<\alpha$	 \\ \hline 
Total	&	$<\alpha$	&	$<\alpha$	&	$<\alpha$	&	$<\alpha$	&	$<\alpha$	&	$<\alpha$	&	$<\alpha$	&	$<\alpha$	&	$<\alpha$	&	$<\alpha$	&	$<\alpha$	&	$<\alpha$	&	$<\alpha$	 \\ \hline 
 
\end{tabular}
\caption{Computed p-values relative to the null hypothesis that the number of translated male pronouns is not significantly greater than that of female pronouns, organized for each language and each occupation category. Cells corresponding to the acceptance of the null hypothesis are marked in blue, and within those cells, those corresponding to cases in which the complementary null hypothesis (that the number of female pronouns is not significantly greater than that of male pronouns) was rejected are marked with a darker shade of the same color. A significance level of $\alpha =.05$ was adopted. Asterisks indicate cases in which all pronouns are translated with gender neutral pronouns.}
\label{tab:pvalues-MF}
\end{table}

\begin{table}[H]
\centering
\footnotesize\setlength{\tabcolsep}{2.5pt}
\small
\begin{tabular}{|m{1.75cm}|cccccccccccc|c|}
\hline
& Mal. & Est. & Fin. & Hun. & Arm. & Ben. & Jap. & Tur. & Yor. & Bas. & Swa. & Chi. & Total \\ \hline
Service	&	$<\alpha$	&	$<\alpha$	&	$<\alpha$	&	$<\alpha$	&	$<\alpha$	&	$<\alpha$	&	$<\alpha$	&	$<\alpha$	&	$<\alpha$	&	\cellcolor{blue!45}$1.0$	&	$<\alpha$	&	$<\alpha$	&	$<\alpha$	 \\ \hline 
STEM	&	$<\alpha$	&	$<\alpha$	&	$<\alpha$	&	$<\alpha$	&	$<\alpha$	&	$<\alpha$	&	$<\alpha$	&	$<\alpha$	&	$<\alpha$	&	\cellcolor{blue!45}$.984$	&	$<\alpha$	&	\cellcolor{blue!25}$.07$	&	$<\alpha$	 \\ \hline 
Farming 
 Fishing 
 Forestry	&	$<\alpha$	&	$<\alpha$	&	$<\alpha$	&	$<\alpha$	&	$<\alpha$	&	\cellcolor{blue!25}$.135$	&	$<\alpha$	&	$<\alpha$	&	\cellcolor{blue!25}$.068$	&	\cellcolor{blue!45}$1.0$	&	$<\alpha$	&	$<\alpha$	&	$<\alpha$	 \\ \hline 
Corporate	&	$<\alpha$	&	$<\alpha$	&	$<\alpha$	&	$<\alpha$	&	$<\alpha$	&	$<\alpha$	&	$<\alpha$	&	$<\alpha$	&	$<\alpha$	&	\cellcolor{blue!45}$1.0$	&	$<\alpha$	&	$<\alpha$	&	$<\alpha$	 \\ \hline 
Healthcare	&	$<\alpha$	&	$<\alpha$	&	$<\alpha$	&	$<\alpha$	&	$<\alpha$	&	$<\alpha$	&	$<\alpha$	&	$<\alpha$	&	\cellcolor{blue!25}$.39$	&	\cellcolor{blue!45}$1.0$	&	$<\alpha$	&	\cellcolor{blue!25}$.088$	&	$<\alpha$	 \\ \hline 
Legal	&	$<\alpha$	&	$<\alpha$	&	$<\alpha$	&	$<\alpha$	&	$<\alpha$	&	$<\alpha$	&	\cellcolor{blue!25}$.145$	&	$<\alpha$	&	$<\alpha$	&	\cellcolor{blue!25}$.771$	&	$<\alpha$	&	$<\alpha$	&	$<\alpha$	 \\ \hline 
Arts 
 Entertainment	&	$<\alpha$	&	$<\alpha$	&	$<\alpha$	&	$<\alpha$	&	$<\alpha$	&	\cellcolor{blue!25}$.07$	&	$<\alpha$	&	$<\alpha$	&	$<\alpha$	&	\cellcolor{blue!45}$1.0$	&	$<\alpha$	&	$<\alpha$	&	$<\alpha$	 \\ \hline 
Education	&	$<\alpha$	&	$<\alpha$	&	$<\alpha$	&	$<\alpha$	&	$<\alpha$	&	$<\alpha$	&	\cellcolor{blue!25}$.093$	&	$<\alpha$	&	$<\alpha$	&	\cellcolor{blue!25}$.5$	&	$<\alpha$	&	\cellcolor{blue!25}$.068$	&	$<\alpha$	 \\ \hline 
Production	&	$<\alpha$	&	$<\alpha$	&	$<\alpha$	&	$<\alpha$	&	$<\alpha$	&	$<\alpha$	&	$<\alpha$	&	\cellcolor{blue!25}$.412$	&	\cellcolor{blue!45}$1.0$	&	\cellcolor{blue!45}$1.0$	&	$<\alpha$	&	$<\alpha$	&	$<\alpha$	 \\ \hline 
Construction 
 Extraction	&	$<\alpha$	&	$<\alpha$	&	$<\alpha$	&	$<\alpha$	&	$<\alpha$	&	$<\alpha$	&	$<\alpha$	&	$<\alpha$	&	\cellcolor{blue!25}$.92$	&	\cellcolor{blue!45}$1.0$	&	$<\alpha$	&	$<\alpha$	&	$<\alpha$	 \\ \hline 
Total	&	$<\alpha$	&	$<\alpha$	&	$<\alpha$	&	$<\alpha$	&	$<\alpha$	&	$<\alpha$	&	$<\alpha$	&	$<\alpha$	&	$<\alpha$	&	\cellcolor{blue!45}$1.0$	&	$<\alpha$	&	$<\alpha$	&	$<\alpha$	 \\ \hline 

\end{tabular}
\caption{Computed p-values relative to the null hypothesis that the number of translated male pronouns is not significantly greater than that of gender neutral pronouns, organized for each language and each occupation category. Cells corresponding to the acceptance of the null hypothesis are marked in blue, and within those cells, those corresponding to cases in which the complementary null hypothesis (that the number of gender neutral pronouns is not significantly greater than that of male pronouns) was rejected are marked with a darker shade of the same color. A significance level of $\alpha =.05$ was adopted. Asterisks indicate cases in which all pronouns are translated with gender neutral pronouns.}
\label{tab:pvalues-MN}
\end{table}

\begin{table}[H]
\centering
\footnotesize\setlength{\tabcolsep}{2.5pt}
\small
\begin{tabular}{|m{1.75cm}|cccccccccccc|c|}
\hline
& Mal. & Est. & Fin. & Hun. & Arm. & Ben. & Jap. & Tur. & Yor. & Bas. & Swa. & Chi. & Total \\ \hline
Service	&	\cellcolor{blue!45}$1.0$	&	\cellcolor{blue!45}$1.0$	&	\cellcolor{blue!45}$1.0$	&	\cellcolor{blue!45}$1.0$	&	\cellcolor{blue!45}$.981$	&	$<\alpha$	&	$<\alpha$	&	$<\alpha$	&	$<\alpha$	&	$<\alpha$	&	\cellcolor{blue!45}$1.0$	&	$<\alpha$	&	$<\alpha$	 \\ \hline 
STEM	&	\cellcolor{blue!25}$.84$	&	\cellcolor{blue!45}$.978$	&	\cellcolor{blue!45}$.998$	&	\cellcolor{blue!45}$.993$	&	\cellcolor{blue!25}$.84$	&	$<\alpha$	&	$<\alpha$	&	\cellcolor{blue!25}$.079$	&	$<\alpha$	&	$<\alpha$	&	\cellcolor{blue!25}$.84$	&	$<\alpha$	&	$<\alpha$	 \\ \hline 
Farming 
 Fishing 
 Forestry	&	\cellcolor{blue!25}$*$	&	\cellcolor{blue!25}$*$	&	\cellcolor{blue!45}$.999$	&	\cellcolor{blue!45}$1.0$	&	\cellcolor{blue!25}$*$	&	\cellcolor{blue!25}$.167$	&	\cellcolor{blue!25}$.169$	&	\cellcolor{blue!25}$.292$	&	$<\alpha$	&	$<\alpha$	&	\cellcolor{blue!25}$*$	&	\cellcolor{blue!25}$.083$	&	\cellcolor{blue!25}$.147$	 \\ \hline 
Corporate	&	\cellcolor{blue!25}$*$	&	\cellcolor{blue!45}$1.0$	&	\cellcolor{blue!45}$1.0$	&	\cellcolor{blue!45}$1.0$	&	\cellcolor{blue!45}$.996$	&	$<\alpha$	&	$<\alpha$	&	$<\alpha$	&	$<\alpha$	&	$<\alpha$	&	\cellcolor{blue!45}$.977$	&	$<\alpha$	&	$<\alpha$	 \\ \hline 
Healthcare	&	\cellcolor{blue!45}$1.0$	&	\cellcolor{blue!45}$1.0$	&	\cellcolor{blue!45}$1.0$	&	\cellcolor{blue!45}$1.0$	&	\cellcolor{blue!45}$1.0$	&	\cellcolor{blue!25}$.086$	&	$<\alpha$	&	\cellcolor{blue!25}$.87$	&	$<\alpha$	&	$<\alpha$	&	\cellcolor{blue!45}$1.0$	&	$<\alpha$	&	\cellcolor{blue!45}$.977$	 \\ \hline 
Legal	&	\cellcolor{blue!25}$*$	&	\cellcolor{blue!45}$.961$	&	\cellcolor{blue!45}$.985$	&	\cellcolor{blue!45}$.961$	&	\cellcolor{blue!25}$*$	&	$<\alpha$	&	\cellcolor{blue!25}$.086$	&	\cellcolor{blue!25}$*$	&	\cellcolor{blue!25}$.178$	&	$<\alpha$	&	\cellcolor{blue!25}$*$	&	\cellcolor{blue!25}$*$	&	\cellcolor{blue!25}$.072$	 \\ \hline 
Arts 
 Entertainment	&	\cellcolor{blue!25}$.92$	&	\cellcolor{blue!45}$.994$	&	\cellcolor{blue!45}$.999$	&	\cellcolor{blue!45}$.998$	&	\cellcolor{blue!45}$.998$	&	\cellcolor{blue!25}$.067$	&	$<\alpha$	&	$<\alpha$	&	$<\alpha$	&	$<\alpha$	&	\cellcolor{blue!25}$.92$	&	\cellcolor{blue!25}$.162$	&	\cellcolor{blue!25}$.097$	 \\ \hline 
Education	&	\cellcolor{blue!25}$*$	&	\cellcolor{blue!45}$1.0$	&	\cellcolor{blue!45}$.999$	&	\cellcolor{blue!45}$.999$	&	\cellcolor{blue!45}$1.0$	&	\cellcolor{blue!25}$.058$	&	$<\alpha$	&	\cellcolor{blue!45}$1.0$	&	\cellcolor{blue!25}$.164$	&	\cellcolor{blue!25}$.052$	&	\cellcolor{blue!45}$.995$	&	\cellcolor{blue!25}$.052$	&	\cellcolor{blue!45}$.992$	 \\ \hline 
Production	&	\cellcolor{blue!45}$.996$	&	\cellcolor{blue!45}$1.0$	&	\cellcolor{blue!45}$1.0$	&	\cellcolor{blue!45}$1.0$	&	\cellcolor{blue!45}$1.0$	&	\cellcolor{blue!25}$.113$	&	$<\alpha$	&	$<\alpha$	&	$<\alpha$	&	$<\alpha$	&	\cellcolor{blue!45}$1.0$	&	$<\alpha$	&	$<\alpha$	 \\ \hline 
Construction 
 Extraction	&	\cellcolor{blue!25}$.84$	&	\cellcolor{blue!45}$.996$	&	\cellcolor{blue!45}$1.0$	&	\cellcolor{blue!45}$1.0$	&	\cellcolor{blue!25}$*$	&	$<\alpha$	&	$<\alpha$	&	$<\alpha$	&	$<\alpha$	&	$<\alpha$	&	\cellcolor{blue!45}$1.0$	&	$<\alpha$	&	$<\alpha$	 \\ \hline 
Total	&	\cellcolor{blue!45}$1.0$	&	\cellcolor{blue!45}$1.0$	&	\cellcolor{blue!45}$1.0$	&	\cellcolor{blue!45}$1.0$	&	\cellcolor{blue!45}$1.0$	&	$<\alpha$	&	$<\alpha$	&	$<\alpha$	&	$<\alpha$	&	$<\alpha$	&	\cellcolor{blue!45}$1.0$	&	$<\alpha$	&	$<\alpha$	 \\ \hline 

\end{tabular}
\caption{Computed p-values relative to the null hypothesis that the number of translated gender neutral pronouns is not significantly greater than that of female pronouns, organized for each language and each occupation category. Cells corresponding to the acceptance of the null hypothesis are marked in blue, and within those cells, those corresponding to cases in which the complementary null hypothesis (that the number of female pronouns is not significantly greater than that of gender neutral pronouns) was rejected are marked with a darker shade of the same color. A significance level of $\alpha =.05$ was adopted. Asterisks indicate cases in which all pronouns are translated with gender neutral pronouns.}
\label{tab:pvalues-NF}
\end{table}

\section{Distribution of translated gender pronouns per language}

\begin{figure}[H]
	\centering
	\includegraphics[width=10cm]{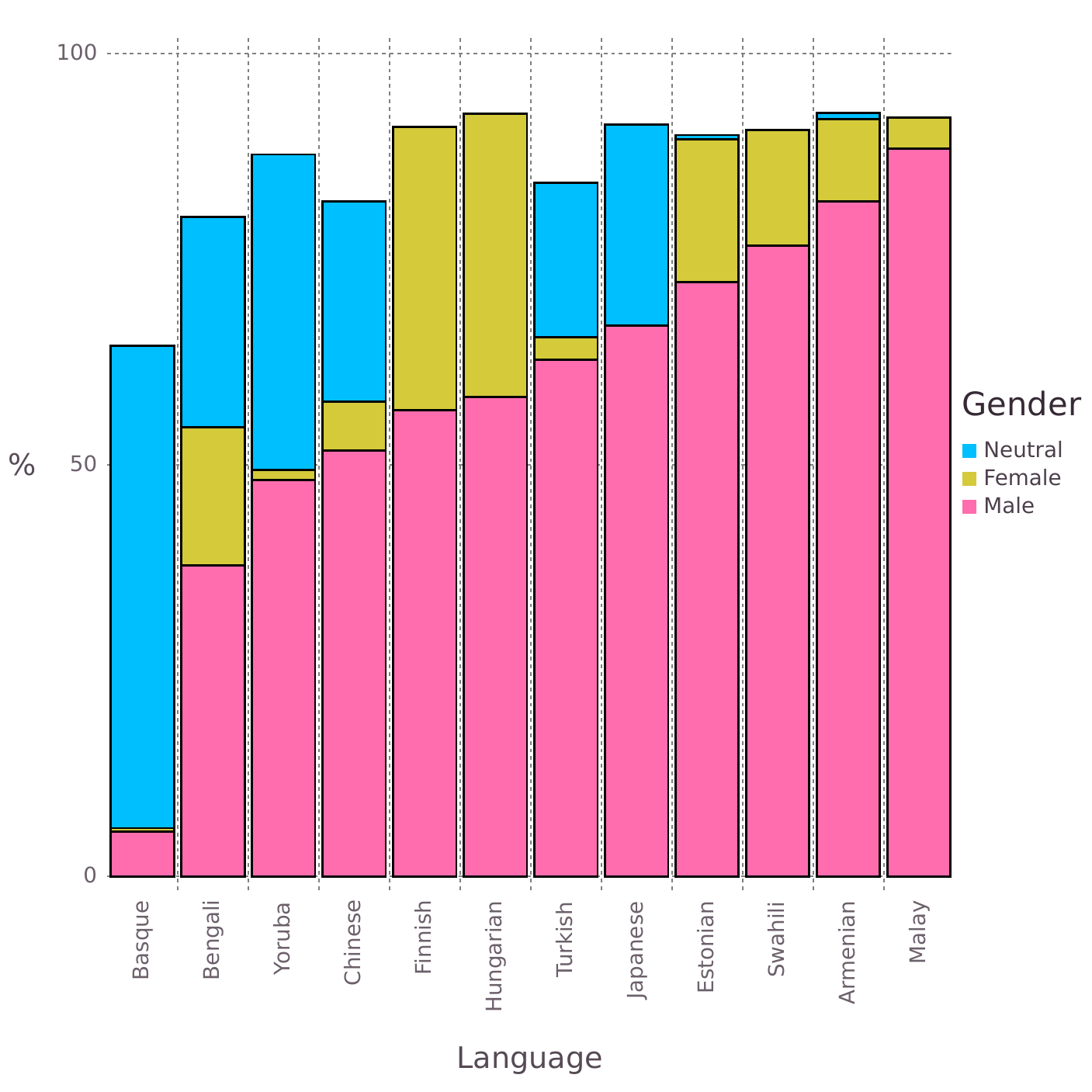}
	\caption{The distribution of pronominal genders per language also suggests a tendency towards male defaults, with female pronouns reaching as low as $0.196\%$ and $1.865\%$ for Japanese and Chinese respectively. Once again not all bars add up to $100\%$ , as there is a fair amount of translated sentences for which we cannot obtain a gender pronoun, particularly in Basque. Among all tested languages, Basque was the only one to yield more gender neutral than male pronouns, with Bengali and Yoruba following after in this order. Languages are sorted with respect to the proportions of male, female and neutral translated pronouns respectively.} 
	\label{fig:gender-by-language}
\end{figure}

We have taken the care of experimenting with a fair amount of different gender neutral languages. Because of that, another sensible way of coalescing our data is by language groups, as shown in Table \ref{tab:gender-by-language}. This can help us visualize the effect of different cultures in the genesis -- or lack thereof -- of gender bias. Nevertheless, the barplots in Figure \ref{fig:gender-by-language} are perhaps most useful to identifying the difficulty of extracting a gender pronoun when translating from certain languages. Basque is a good example of this difficulty, although the quality of Bengali, Yoruba, Chinese and Turkish translations are also compromised.

\begin{table}[H]
\small{
	\centering
	\begin{tabular}{|c|c|c|c|}
	\hline
	Language 	& Female ($\%$) 	& Male ($\%$)		& Neutral ($\%$)	\\ \hline
	\hline
	Malay & 3.827 & 88.420 & 0.000 \\ \hline
	Estonian & 17.370 & 72.228 & 0.491 \\ \hline
	Finnish & 34.446 & 56.624 & 0.000 \\ \hline
	Hungarian & 34.347 & 58.292 & 0.000 \\ \hline
	Armenian & 10.010 & 82.041 & 0.687 \\ \hline
	Bengali & 16.765 & 37.782 & 25.63 \\ \hline
	Japanese & 0.000 & 66.928 & 24.436 \\ \hline
	Turkish & 2.748 & 62.807 & 18.744 \\ \hline
	Yoruba & 1.178 & 48.184 & 38.371 \\ \hline
	Basque & 0.393 & 5.496 & 58.587 \\ \hline
	Swahili & 14.033 & 76.644 & 0.000 \\ \hline
	Chinese & 5.986 & 51.717 & 24.338 \\ \hline
	\end{tabular}
	\caption{Percentage of female, male and neutral gender pronouns obtained for each language, averaged over all occupations detailed in Table} \ref{tab:gender-neutral-languages}. Note that rows do not in general add up to $100\%$, as there is a fair amount of translated sentences for which we cannot obtain a gender pronoun.
	\label{tab:gender-by-language}
	}
\end{table}

\section{Distribution of translated gender pronouns for varied adjectives}

We queried the 1000 most frequently used adjectives in English, as classified in the COCA corpus [https://corpus.byu.edu/coca/], but since not all of them were readily applicable to the sentence template we used, we filtered the N adjectives that would fit the templates and made sense for describing a human being. The list of adjectives extracted from the corpus is available on the Github repository: https://github.com/marceloprates/Gender-Bias.

Apart from occupations, which we have exhaustively examined by collecting labor data from the U.S. Bureau of Labor Statistics, we have also selected a small subset of adjectives from the Corpus of Contemporary American English (COCA) https://corpus.byu.edu/coca/, in an attempt to provide preliminary evidence that the phenomenon of gender bias may extend beyond the professional context examined in this paper. Because a large number of adjectives are not applicable to human subjects, we manually curated a reasonable subset of such words. The template used for adjectives is similar to that used for occupations, and is provided again for reference in Table \ref{tab:templates}.

Once again the data points towards male defaults, but some variation can be observed throughout different adjectives. Sentences containing the words \emph{Shy}, \emph{Attractive}, \emph{Happy}, \emph{Kind} and \emph{Ashamed} are predominantly female translated (\emph{Attractive} is translated as female and gender-neutral in equal parts), while \emph{Arrogant}, \emph{Cruel} and \emph{Guilty} are disproportionately translated with male pronouns (\emph{Guilty} is in fact never translated with female or neutral pronouns).

\begin{table}[H]
\small{
	\centering
	\begin{tabular}{|c|c|c|c|}
	\hline
	Adjective & Female ($\%$) & Male ($\%$) & Neutral ($\%$) \\ \hline \hline
	Happy & 36.364 & 27.273 & 18.182 \\ \hline
	Sad & 18.182 & 36.364 & 18.182 \\ \hline
	Right & 0.000 & 63.636 & 27.273 \\ \hline
	Wrong & 0.000 & 54.545 & 36.364 \\ \hline
	Afraid & 9.091 & 54.545 & 0.000 \\ \hline
	Brave & 9.091 & 63.636 & 18.182 \\ \hline
	Smart & 18.182 & 45.455 & 18.182 \\ \hline
	Dumb & 18.182 & 36.364 & 18.182 \\ \hline
	Proud & 9.091 & 72.727 & 9.091 \\ \hline
	Strong & 9.091 & 54.545 & 18.182 \\ \hline
	Polite & 18.182 & 45.455 & 18.182 \\ \hline
	Cruel & 9.091 & 63.636 & 18.182 \\ \hline
	Desirable & 9.091 & 36.364 & 45.455 \\ \hline
	Loving & 18.182 & 45.455 & 27.273 \\ \hline
	Sympathetic & 18.182 & 45.455 & 18.182 \\ \hline
	Modest & 18.182 & 45.455 & 27.273 \\ \hline
	Successful & 9.091 & 54.545 & 27.273 \\ \hline
	Guilty & 0.000 & 72.727 & 0.000 \\ \hline
	Innocent & 9.091 & 54.545 & 9.091 \\ \hline
	Mature & 36.364 & 36.364 & 9.091 \\ \hline
	Shy & 36.364 & 27.273 & 27.273 \\ \hline \hline
	Total & 30.3 & 98.1 & 41.7 \\ \hline

	\end{tabular}
	\caption{Number of female, male and neutral pronominal genders in the translated sentences for each selected adjective.}
	\label{tab:gender-by-adjective}
	}
\end{table}

\begin{figure}[H]
	\centering
	\includegraphics[width=10cm]{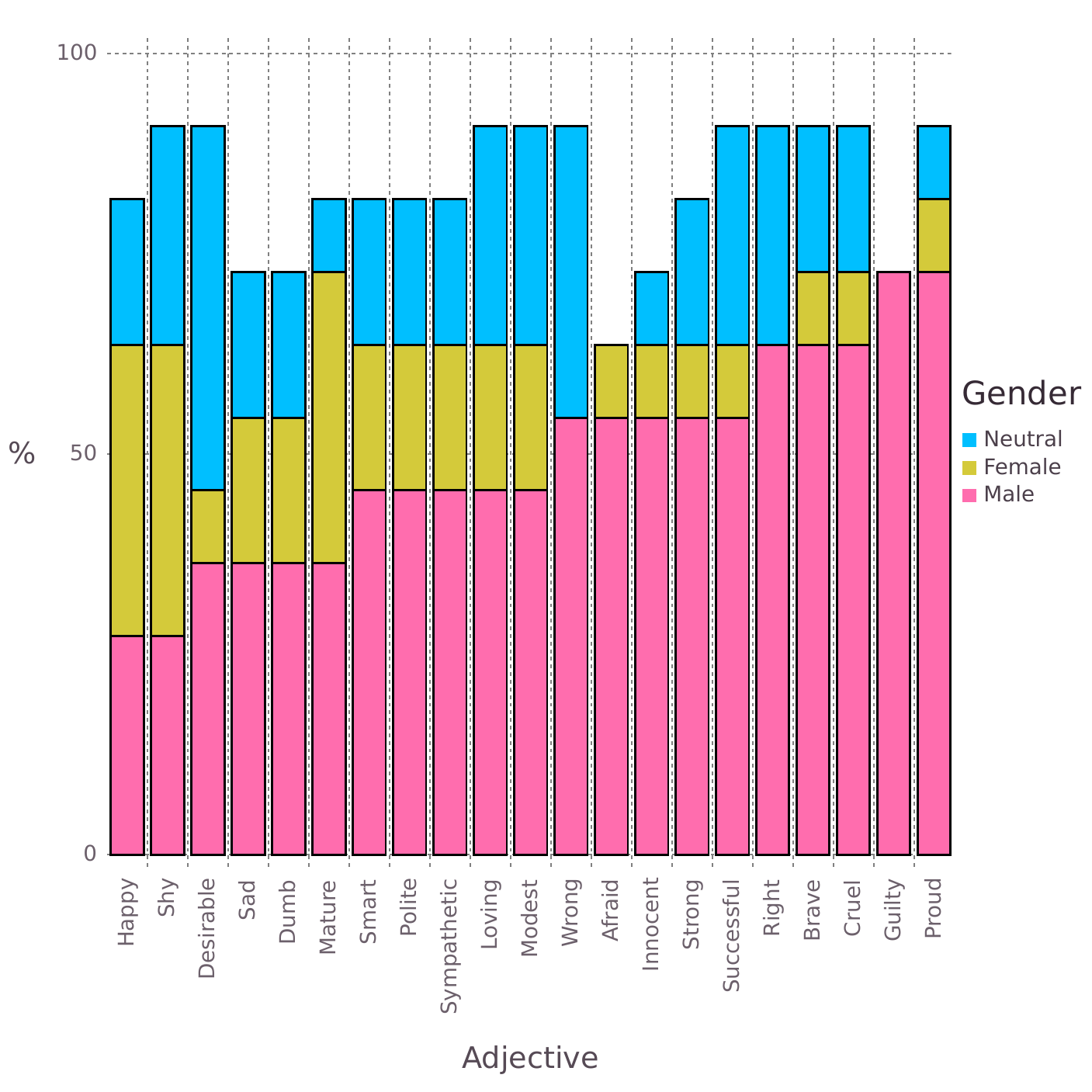}
	\caption{The distribution of pronominal genders for each word in Table} \ref{tab:adjectives} shows how stereotypical gender roles can play a part on the automatic translation of simple adjectives. One can see that adjectives such as \emph{Shy} and \emph{Desirable}, \emph{Sad} and \emph{Dumb} amass at the female side of the spectrum, contrasting with \emph{Proud}, \emph{Guilty}, \emph{Cruel} and \emph{Brave} which are almost exclusively translated with male pronouns. 
	\label{fig:barplot-adjectives}
\end{figure}

\section{Comparison with women participation data across job positions}\label{sec:comparison-women-participation}

A sensible objection to the conclusions we draw from our study is that the perceived gender bias in Google Translate results stems from the fact that possibly female participation in some job positions is itself low. We must account for the possibility that the statistics of gender pronouns in Google Translate outputs merely reflects the demographics of male-dominated fields (male-dominated fields can be considered those that have less than 25\% of women participation\cite{WB2014}, according to the U.S. Department of Labor Women's Bureau). In this context, the argument in favor of a critical revision of statistic translation algorithms weakens considerably, and possibly shifts the blame away from these tools.

\begin{figure}[H]
	\centering
	\includegraphics[width=10cm]{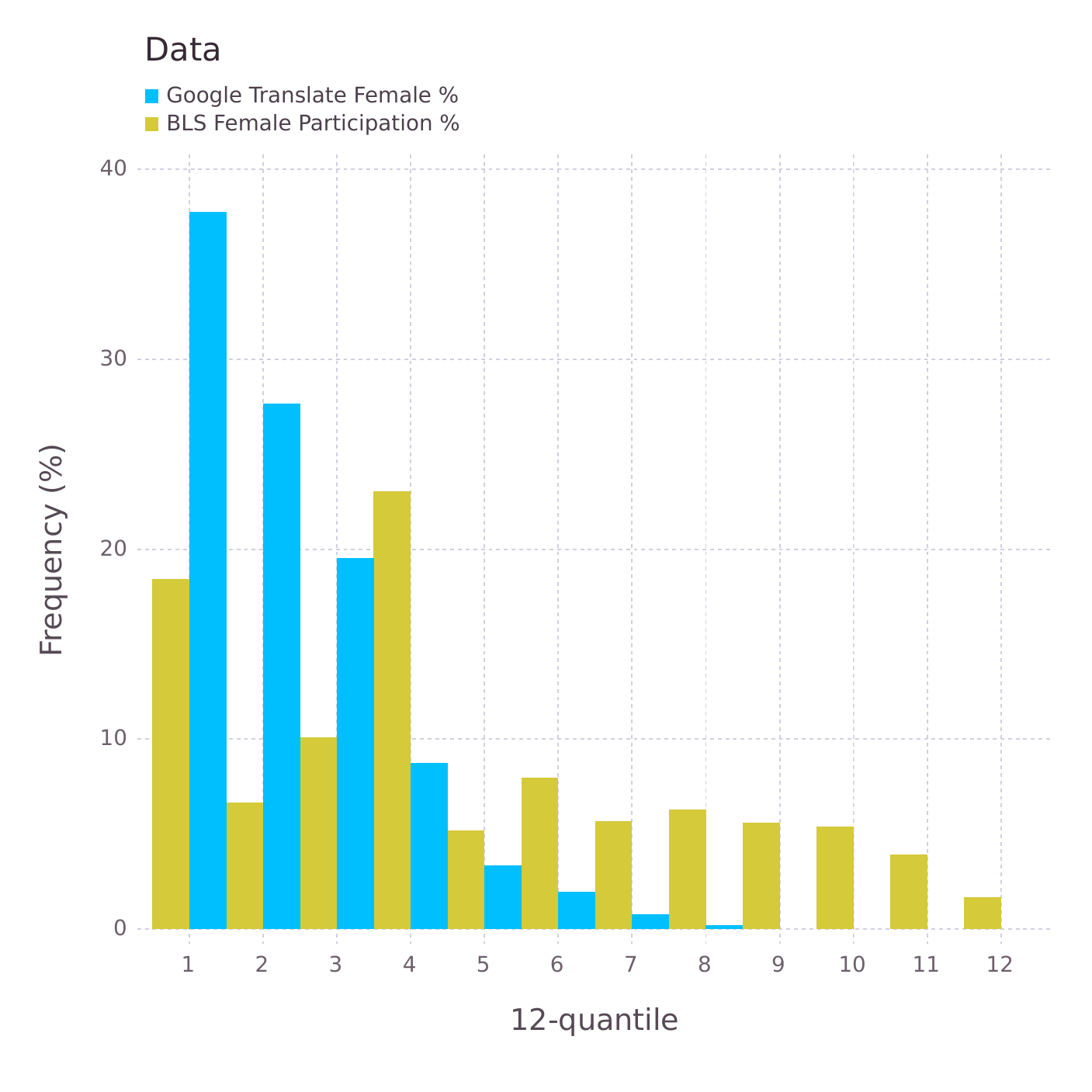}
	\caption{Women participation ($\%$) data obtained from the U.S. Bureau of Labor Statistics allows us to assess whether the Google Translate bias towards male defaults is at least to some extent explained by small frequencies of female workers in some job positions. Our data does not make a very good case for that hypothesis: the total frequency of translated female pronouns (in blue) for each 12-quantile does not seem to respond to the higher proportion of female workers (in yellow) in the last quantiles.}
	\label{fig:histogram-compare-gt-real}
\end{figure}

The U.S. Bureau of Labor Statistics data summarized in Table \ref{tab:occupations} contains statistics about the percentage of women participation in each occupation category. This data is also available for each individual occupation, which allows us to compute the frequency of women participation for each 12-quantile. We carried the same computation in the context of frequencies of translated female pronouns, and the resulting histograms are plotted side-by-side in Figure \ref{fig:histogram-compare-gt-real}. The data shows us that Google Translate outputs fail to follow the real-world distribution of female workers across a comprehensive set of job positions. The distribution of translated female pronouns is consistently inversely distributed, with female pronouns accumulating in the first 12-quantile. By contrast, BLS data shows that female participation peaks in the fourth 12-quantile and remains significant throughout the next ones.

Averaged over occupations and languages, sentences are translated with female pronouns $11.76\%$ of the time. In contrast, the gender participation frequency for female workers averaged over all occupations in the BLS report yields a consistently larger figure of $35.94\%$. The variance reported for the translation results is also lower, at $\approx 0.028$ in contrast with the report's $\approx 0.067$. We ran an one-sided t-test to evaluate the null hypothesis that the female participation frequency is not significantly greater then the GT female pronoun frequency for the same job positions, obtaining a p-value $p \approx 6.2 10^{-94}$ vastly inferior to our confidence level of $\alpha = 0.005$ and thus rejecting H0 and concluding that Google Translate's female translation frequencies sub-estimates female participation frequencies in US job positions. As a result, it is not possible to understand this asymmetry as a reflection of workplace demographics, and the prominence of male defaults in Google Translate is, we believe, yet lacking a clear justification.

\section{Discussion}

\begin{figure}[h]
    \centering
    \includegraphics[width=\linewidth]{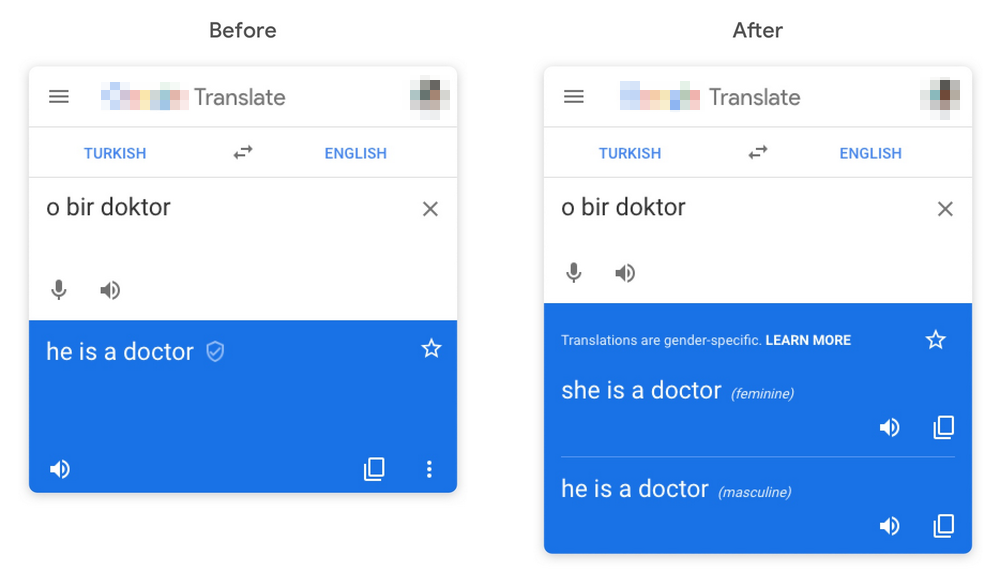}
    \caption{Comparison between the GUI of Google Translate before (left) and after (right) the introduction of the new feature intended to promote gender fairness in translation. The results described in this paper relate to the older version.}
    \label{fig:gender-before-after}
\end{figure}

At the time of the writing up this paper, Google Translate offered only one official translation for each input word, along with a list of synonyms. In this context, all experiments reported here offer an analysis of a   ``screenshot'' of that tool as of August 2018, the moment they were carried out. A preprint version of this paper was posted the in well-known Cornell University-based \url{arXiv.org} open repository on September 6, 2018. The manuscript soon enjoyed a significant amount of media coverage, featuring on \emph{The Register} \cite{clauburn2018boffins}, \emph{Datanews} \cite{bellens2018sexiste}, \emph{t3n} \cite{rixecker2018sexistische}, among others, and more recently on \emph{Slator} \cite{Dino2019hesaid} and \emph{Jornal do Comercio} \cite{knebel2019nosrobos}. On December 6, 2018 the company's policy changed, and a statement was released detailing their efforts to reduce gender bias on Google Translate, which included a new feature presenting the user with a feminine as well as a masculine official translation (Figure \ref{fig:gender-before-after}). According to the company, this decision is part of a broader goal of promoting fairness and reducing biases in machine learning. They also acknowledged the technical reasons behind gender bias in their model, stating that:

\begin{quote}
    \textsl{Google Translate learns from hundreds of millions of already-translated examples from the web. Historically, it has provided only one translation for a query, even if the translation could have either a feminine or masculine form. So when the model produced one translation, it inadvertently replicated gender biases that already existed. For example: it would skew masculine for words like ``strong'' or ``doctor,'' and feminine for other words, like ``nurse'' or ``beautiful.''}
\end{quote}

Their statement is very similar to the conclusions drawn on this paper, as is their motivation for redesigning the tool. As authors, we are incredibly happy to see our vision and beliefs align with those of Google in such a short timespan from the initial publishing of our work, although the company's statement does not cite any study or report in particular and thus we cannot know for sure whether this paper had an effect on their decision or not. Regardless of whether their decision was monocratic, guided by public opinion or based on published research, we understand it as an important first step on an ongoing fight against algorithmic bias, and we praise the Google Translate team for their efforts.

Google Translate's new feminine and masculine forms for translated sentences exemplifies how, as this paper also suggests, machine learning translation tools can be \emph{debiased}, dropping the need for resorting to a balanced training set. However, it should be noted that important as it is, GT's new feature is still a first step. It does not address all of the shortcomings described in this paper, and the limited language coverage means that many users will still experience gender biased translation results. Furthermore, the system does not yet have support for non-binary results, which may exclude part of their user base.

In addition, one should note that further evidence is mounting about the kind of bias examined in this paper: it is becoming clear that this is a statistical phenomenon independent from any proprietary tool. In this context, the research carried out in \cite{bolukbasi2016man} presents a very convincing argument for the sensitivity of word embeddings to gender bias in the training dataset. This suggests that machine translation engineers should be especially aware of their training data when designing a system. It is not feasible to train these models on unbiased texts, as they are probably scarce. What must be done instead is to engineer solutions to remove bias from the system after an initial training, which seems to be the goal of Google Translate's recent efforts. Fortunately, as \cite{bolukbasi2016man} also show, debiasing can be implemented with relatively low effort and modest resources. The technology to promote social justice on machine translation in particular and machine learning in general is often already available. The most significant effort which must be taken in this context is to promote social awareness on these issues so that society can be invited into the conversation.

\section{Conclusions}

In this paper, we have provided evidence that statistical translation tools such as Google Translate can exhibit gender biases and a strong tendency towards male defaults. Although implicit, these biases possibly stem from the real world data which is used to train them, and in this context possibly provide a window into the way our society talks (and writes) about women in the workplace. In this paper, we suggest that and test the hypothesis that statistical translation tools can be probed to yield insights about stereotypical gender roles in our society -- or at least in their training data. By translating professional-related sentences such as ``He/She is an engineer'' from gender neutral languages such as Hungarian and Chinese into English, we were able to collect statistics about the asymmetry between female and male pronominal genders in the translation outputs. Our results show that male defaults are not only prominent, but exaggerated in fields suggested to be troubled with gender stereotypes, such as STEM (Science, Technology, Engineering and Mathematics) occupations. And because Google Translate typically uses English as a \emph{lingua franca} to translate between other languages (e.g. Chinese $\rightarrow$ English $\rightarrow$ Portuguese) \cite{LanguageSupportNMT2018,boitet2010mt}, our findings possibly extend to translations between gender neutral languages and non-gender neutral languages (apart from English) in general, although we have not tested this hypothesis.

Our results seem to suggest that this phenomenon extends beyond the scope of the workplace, with the proportion of female pronouns varying significantly according to adjectives used to describe a person. Adjectives such as \emph{Shy} and \emph{Desirable} are translated with a larger proportion of female pronouns, while \emph{Guilty} and \emph{Cruel} are almost exclusively translated with male ones. Different languages also seemingly have a significant impact in machine gender bias, with Hungarian exhibiting a better equilibrium between male and female pronouns than, for instance, Chinese. Some languages such as Yoruba and Basque were found to translate sentences with gender neutral pronouns very often, although this is the exception rather than the rule and Basque also exhibits a high frequency of phrases for which we could not automatically extract a gender pronoun.

In order to strengthen our results, we ran pronominal gender translation statistics against the U.S. Bureau of Labor Statistics data on the frequency of women participation for each job position. Although Google Translate exhibits male defaults, this phenomenon may merely reflect the unequal distribution of male and female workers in some job positions. To test this hypothesis, we compared the distribution of female workers with the frequency of female translations, finding no correlation between said variables. Our data shows that Google Translate outputs fail to reflect the real-world distribution of female workers, under-estimating the expected frequency. That is to say that even if we do not expect a 50:50 distribution of translated gender pronouns, Google Translate exhibits male defaults in a greater frequency that job occupation data alone would suggest. The prominence of male defaults in Google Translate is therefore to the best of our knowledge yet lacking a clear justification.

We think this work sheds new light on a  pressing ethical difficulty arising from modern statistical machine translation, and hope that it will lead to discussions about the role of AI engineers on minimizing potential harmful effects of the current concerns about machine bias. We are optimistic that unbiased results can be obtained with relatively little effort and marginal cost to the performance of current methods, to which current \emph{debiasing} algorithms in the scientific literature are a testament.

\section{Acknowledgments}

This study was financed in part by the Coordenação de Aperfeiçoamento de Pessoal de Nível Superior - Brasil (CAPES) - Finance Code 001 and the Conselho Nacional de Desenvolvimento Científico e Tecnológico (CNPq).

This is a pre-print of an article published in Neural Computing and Applications.

\vskip 0.2in

\bibliography{main}

\begin{thebibliography}{10}
\providecommand{\url}[1]{{#1}}
\providecommand{\urlprefix}{URL }
\expandafter\ifx\csname urlstyle\endcsname\relax
  \providecommand{\doi}[1]{DOI~\discretionary{}{}{}#1}\else
  \providecommand{\doi}{DOI~\discretionary{}{}{}\begingroup
  \urlstyle{rm}\Url}\fi

\bibitem{angwin2016machine}
Angwin, J., Larson, J., Mattu, S., Kirchner, L.: Machine bias: {T}here's
  software used across the country to predict future criminals and it's biased
  against blacks (2016).
\newblock \urlprefix\url{https://www.
  propublica.org/article/machine-bias-risk-assessments-in-criminal-sentencing}.
\newblock Last visited 2017-12-17

\bibitem{Bahdanau2014}
Bahdanau, D., Cho, K., Bengio, Y.: Neural machine translation by jointly
  learning to align and translate.
\newblock CoRR \textbf{abs/1409.0473} (2014).
\newblock \urlprefix\url{http://arxiv.org/abs/1409.0473}

\bibitem{bellens2018sexiste}
Bellens, E.: Google translate est sexiste (2018).
\newblock
  \urlprefix\url{https://datanews.levif.be/ict/actualite/google-translate-est-sexiste/article-normal-889277.html?cookie_check=1549374652}.
\newblock [Online; posted 11-September-2018]

\bibitem{boitet2010mt}
Boitet, C., Blanchon, H., Seligman, M., Bellynck, V.: Mt on and for the web.
\newblock In: Natural Language Processing and Knowledge Engineering (NLP-KE),
  2010 International Conference on, pp. 1--10. IEEE (2010)

\bibitem{bolukbasi2016man}
Bolukbasi, T., Chang, K.W., Zou, J.Y., Saligrama, V., Kalai, A.T.: Man is to
  computer programmer as woman is to homemaker? {Debiasing} word embeddings.
\newblock In: Advances in Neural Information Processing Systems, pp. 4349--4357
  (2016)

\bibitem{boroditsky2003sex}
Boroditsky, L., Schmidt, L.A., Phillips, W.: Sex, syntax, and semantics.
\newblock Language in mind: Advances in the study of language and thought pp.
  61--79 (2003)

\bibitem{BLS2017}
{Bureau of Labor Statistics}: {"Table 11: Employed persons by detailed
  occupation, sex, race, and Hispanic or Latino ethnicity, 2017"}.
\newblock Labor force statistics from the current population survey, United
  States Department of Labor (2017)

\bibitem{carl2003recent}
Carl, M., Way, A.: Recent advances in example-based machine translation,
  vol.~21.
\newblock Springer Science \& Business Media (2003)

\bibitem{Chomsky2011}
Chomsky, N.: The golden age: A look at the original roots of artificial
  intelligence, cognitive science, and neuroscience (partial transcript of an
  interview with {N}. {C}homsky at {MIT150} {S}ymposia: Brains, minds and
  machines symposium (2011).
\newblock \urlprefix\url{https://chomsky.info/20110616/}.
\newblock Last visited 2017-12-26

\bibitem{clauburn2018boffins}
Clauburn, T.: Boffins bash google translate for sexism (2018).
\newblock
  \urlprefix\url{https://www.theregister.co.uk/2018/09/10/boffins_bash_google_translate_for_sexist_language/}.
\newblock [Online; posted 10-September-2018]

\bibitem{dascal1982universal}
Dascal, M.: Universal language schemes in {E}ngland and {F}rance, 1600-1800
  comments on {J}ames {K}nowlson.
\newblock Studia leibnitiana \textbf{14}(1), 98--109 (1982)

\bibitem{Dino2019hesaid}
Diño, G.: He said, she said: Addressing gender in neural machine translation
  (2019).
\newblock
  \urlprefix\url{https://slator.com/technology/he-said-she-said-addressing-gender-in-neural-machine-translation/}.
\newblock [Online; posted 22-January-2019]

\bibitem{wals}
Dryer, M.S., Haspelmath, M. (eds.): WALS Online.
\newblock Max Planck Institute for Evolutionary Anthropology, Leipzig (2013).
\newblock \urlprefix\url{http://wals.info/}

\bibitem{Firat2017}
Firat, O., Cho, K., Sankaran, B., Yarman{-}Vural, F.T., Bengio, Y.: Multi-way,
  multilingual neural machine translation.
\newblock Computer Speech {\&} Language \textbf{45}, 236--252 (2017).
\newblock \doi{10.1016/j.csl.2016.10.006}.
\newblock \urlprefix\url{https://doi.org/10.1016/j.csl.2016.10.006}

\bibitem{garcia2016racist}
Garcia, M.: Racist in the machine: The disturbing implications of algorithmic
  bias.
\newblock World Policy Journal \textbf{33}(4), 111--117 (2016)

\bibitem{LanguageSupportNMT2018}
Google: Language support for the neural machine translation model (2017).
\newblock
  \urlprefix\url{https://cloud.google.com/translate/docs/languages\#languages-nmt}.
\newblock Last visited 2018-3-19

\bibitem{gordin2015scientific}
Gordin, M.D.: Scientific Babel: How science was done before and after global
  English.
\newblock University of Chicago Press (2015)

\bibitem{hajian2016algorithmic}
Hajian, S., Bonchi, F., Castillo, C.: Algorithmic bias: From discrimination
  discovery to fairness-aware data mining.
\newblock In: Proceedings of the 22nd ACM SIGKDD international conference on
  knowledge discovery and data mining, pp. 2125--2126. ACM (2016)

\bibitem{hutchins1986machine}
Hutchins, W.J.: Machine translation: past, present, future.
\newblock Ellis Horwood Chichester (1986)

\bibitem{wu2016google}
Johnson, M., Schuster, M., Le, Q.V., Krikun, M., Wu, Y., Chen, Z., Thorat, N.,
  Vi{\'{e}}gas, F.B., Wattenberg, M., Corrado, G., Hughes, M., Dean, J.:
  Google's multilingual neural machine translation system: Enabling zero-shot
  translation.
\newblock {TACL} \textbf{5}, 339--351 (2017).
\newblock
  \urlprefix\url{https://transacl.org/ojs/index.php/tacl/article/view/1081}

\bibitem{kay1984sapir}
Kay, P., Kempton, W.: What is the sapir-whorf hypothesis?
\newblock American anthropologist \textbf{86}(1), 65--79 (1984)

\bibitem{TranslateCommunity}
Kelman, S.: Translate community: Help us improve google translate! (2014).
\newblock
  \urlprefix\url{https://search.googleblog.com/2014/07/translate-community-help-us-improve.html}.
\newblock Last visited 2018-3-12

\bibitem{kirkpatrick2016battling}
Kirkpatrick, K.: Battling algorithmic bias: how do we ensure algorithms treat
  us fairly?
\newblock Communications of the ACM \textbf{59}(10), 16--17 (2016)

\bibitem{knebel2019nosrobos}
Knebel, P.: Nós, os robôs e a ética dessa relação (2019).
\newblock
  \urlprefix\url{https://www.jornaldocomercio.com/_conteudo/cadernos/empresas_e_negocios/2019/01/665222-nos-os-robos-e-a-etica-dessa-relacao.html}.
\newblock [Online; posted 4-Februrary-2019]

\bibitem{koehn2009statistical}
Koehn, P.: Statistical machine translation.
\newblock Cambridge University Press (2009)

\bibitem{Moses2007}
Koehn, P., Hoang, H., Birch, A., Callison{-}Burch, C., Federico, M., Bertoldi,
  N., Cowan, B., Shen, W., Moran, C., Zens, R., Dyer, C., Bojar, O.,
  Constantin, A., Herbst, E.: Moses: Open source toolkit for statistical
  machine translation.
\newblock In: {ACL} 2007, Proceedings of the 45th Annual Meeting of the
  Association for Computational Linguistics, June 23-30, 2007, Prague, Czech
  Republic (2007).
\newblock \urlprefix\url{http://aclweb.org/anthology/P07-2045}

\bibitem{locke1955machine}
Locke, W.N., Booth, A.D.: Machine translation of languages: fourteen essays.
\newblock Published jointly by Technology Press of the Massachusetts Institute
  of Technology and Wiley, New York (1955)

\bibitem{racistsoapdispenser2017}
Mills, K.A.: '{R}acist' soap dispenser refuses to help dark-skinned man wash
  his hands - but {T}witter blames 'technology' (2017).
\newblock
  \urlprefix\url{http://www.mirror.co.uk/news/world-news/racist-soap-dispenser-refuses-help-11004385}.
\newblock Last visited 2017-12-17

\bibitem{moss2015can}
Moss-Racusin, C.A., Molenda, A.K., Cramer, C.R.: Can evidence impact attitudes?
  public reactions to evidence of gender bias in stem fields.
\newblock Psychology of Women Quarterly \textbf{39}(2), 194--209 (2015)

\bibitem{norvig2017chomsky}
Norvig, P.: On {C}homsky and the two cultures of statistical learning (2017).
\newblock \urlprefix\url{http://norvig.com/chomsky.html}.
\newblock Last visited 2017-12-17

\bibitem{AlgorithmGtranslateSexist2018}
Olson, P.: The algorithm that helped google translate become sexist (2018).
\newblock
  \urlprefix\url{https://www.forbes.com/sites/parmyolson/2018/02/15/the-algorithm-that-helped-google-translate-become-sexist/#1c1122c27daa}.
\newblock Last visited 2018-3-12

\bibitem{womanunlockphone2017}
Papenfuss, M.: Woman in {C}hina says colleague's face was able to unlock her
  i{P}hone {X} (2017).
\newblock
  \urlprefix\url{http://www.huffpostbrasil.com/entry/iphone-face-recognition-double_us_5a332cbce4b0ff955ad17d50}.
\newblock Last visited 2017-12-17

\bibitem{rixecker2018sexistische}
Rixecker, K.: Google translate verstärkt sexistische vorurteile (2018).
\newblock
  \urlprefix\url{https://t3n.de/news/google-translate-verstaerkt-sexistische-vorurteile-1109449/}.
\newblock [Online; posted 11-September-2018]

\bibitem{santacreu2013female}
Santacreu-Vasut, E., Shoham, A., Gay, V.: Do female/male distinctions in
  language matter? {Evidence} from gender political quotas.
\newblock Applied Economics Letters \textbf{20}(5), 495--498 (2013)

\bibitem{schiebinger2014scientific}
Schiebinger, L.: Scientific research must take gender into account.
\newblock Nature \textbf{507}(7490), 9 (2014)

\bibitem{Gtranslate200daily2017}
Shankland, S.: Google translate now serves 200 million people daily (2017).
\newblock
  \urlprefix\url{https://www.cnet.com/news/google-translate-now-serves-200-million-people-daily/}.
\newblock Last visited 2018-3-12

\bibitem{thompson2014linguistic}
Thompson, A.J.: Linguistic relativity: can gendered languages predict sexist
  attitudes?
\newblock Linguistics Department, Montclair State University (2014)

\bibitem{wang2017deep}
Wang, Y., Kosinski, M.: Deep neural networks are more accurate than humans at
  detecting sexual orientation from facial images.
\newblock Journal of Personality and Social Psychology \textbf{114}(2),
  246--257 (2018)

\bibitem{weaver1955translation}
Weaver, W.: Translation.
\newblock In: W.N. Locke, A.D. Booth (eds.) Machine translation of languages,
  vol.~14, pp. 15--23. Cambridge: Technology Press, MIT (1955).
\newblock \urlprefix\url{http://www.mt-archive.info/Weaver-1949.pdf}.
\newblock Last visited 2017-12-17

\bibitem{WB2014}
{Women's Bureau -- United States Department of Labor}: Traditional and
  nontraditional occupations (2017).
\newblock
  \urlprefix\url{https://www.dol.gov/wb/stats/nontra_traditional_occupations.htm}.
\newblock Last visited 2018-05-30

\end{thebibliography}
\bibliographystyle{spmpsci}

\end{document}